\documentclass[a4paper,twoside]{article}
\newcommand{\affil}[1]{$^{\rm #1}$}
\usepackage[authoryear]{natbib}
\usepackage{amssymb}
\bibpunct{(}{)}{;}{a}{}{,}
\usepackage{graphicx}
\date{} 
\title{\large\bf\flushleft The Mopra Southern Galactic Plane CO Survey}
\author{\parbox{\textwidth}{\flushleft
\vspace{-0.5cm}
{\it Michael G. Burton\affil{A,Z}, C. Braiding\affil{A}, C. Glueck\affil{B}, P. Goldsmith\affil{C}, J. Hawkes\affil{D}, D.J. Hollenbach\affil{E},  C. Kulesa\affil{F}, C. L. Martin\affil{G}, J.L. Pineda\affil{C}, G. Rowell\affil{D}, R. Simon\affil{B}, A.A. Stark\affil{H}, J. Stutzki\affil{B}, N.J.H. Tothill\affil{A,I}, J.S. Urquhart\affil{J}, C. Walker\affil{F}, A.J. Walsh\affil{K} and M. Wolfire\affil{L}}\\
\vspace{0.4cm}
{\small \affil{A}\,School of Physics, University of New South Wales, Sydney, NSW 2052, Australia}\\
{\small \affil{B}\,KOSMA, I. Physikalisches Institut, Universit\"at zu K\"oln, Z\"ulpicher Str.\ 77, 50937 K\"oln, Germany}\\
{\small \affil{C}\,Jet Propulsion Laboratory, California Institute of Technology, 4800 Oak Grove Drive, Pasadena, CA 91109-8099, USA}\\
{\small \affil{D}\,School of Chemistry and Physics, University of Adelaide, Adelaide, SA 5005, Australia}\\
{\small \affil{E}\,Carl Sagan Center, SETI Institute, 189 Bernardo Avenue, Mountain View, CA 94043--5203, USA}\\
{\small \affil{F}\,Steward Observatory, The University of Arizona, 933 N. Cherry Ave., Tucson, AZ 85721, USA}\\
{\small \affil{G}\,Dept. of Physics and Astronomy, Oberlin College, 110 N. Professor St., Oberlin, OH 44074, USA}\\
{\small \affil{H}\,Harvard-Smithsonian Center for Astrophysics, 60 Garden Street, Cambridge, MA 02138, USA}\\
{\small \affil{I}\,School of Computing, Engineering \& Mathematics, University of Western Sydney, Locked Bag 1797, Penrith, NSW 2751, Australia}\\
{\small \affil{J}\,Max-Planck-Institut f\"{u}r Radioastronomie, Auf dem H\"{u}gel 69, D-53121 Bonn, Germany}\\
{\small \affil{K}\,International Centre for Radio Astronomy Research, Curtin University, GPO Box U1987, Perth, WA, Australia}\\
{\small \affil{L}\,Astronomy Department, University of Maryland, College Park, MD 20742, USA}\\
{\small \affil{Z}\,Email: m.burton@unsw.edu.au}}}
\begin{document}
\twocolumn[
\begin{changemargin}{.8cm}{.5cm}
\begin{minipage}{.9\textwidth}
\vspace{-1cm}
\maketitle
%
%
\small{\bf Abstract:}  We present the first results from a new carbon monoxide (CO) survey of the southern Galactic plane being conducted with the Mopra radio telescope in Australia.  The $^{12}$CO, $^{13}$CO and C$^{18}$O J=1--0 lines are being mapped over the $l = 305^{\circ} - 345^{\circ}, b = \pm 0.5^{\circ}$ portion of the 4$^{th}$ quadrant of the Galaxy, at $35''$ spatial and 0.1\,km/s spectral resolution.  The survey is being undertaken with two principal science objectives: (i) to determine where and how molecular clouds are forming in the Galaxy and (ii) to probe the connection between molecular clouds and the ``missing'' gas inferred from gamma-ray observations. We describe the motivation for the survey, the instrumentation and observing techniques being applied, and the data reduction and analysis methodology.    In this paper we present the data from the first degree surveyed, $l = 323^{\circ} - 324^{\circ}, b = \pm 0.5^{\circ}$.    We compare the data to the previous CO survey of this region and present metrics quantifying the performance being achieved; the rms sensitivity per 0.1\,km/s velocity channel is $\sim 1.5$\,K for $\rm ^{12}CO$ and $\sim 0.7$\,K for the other lines.  We also present some results from the region surveyed, including line fluxes, column densities, molecular masses, $\rm ^{12}CO/^{13}CO$ line ratios and $\rm ^{12}CO$ optical depths.  We also examine how these quantities vary as a function of distance from the Sun when averaged over the 1 square degree survey area.   Approximately $\rm 2 \times 10^6\, M_{\odot}$ of molecular gas is found along the G323 sightline, with an average H$_2$ number density of $n_{H_2} \sim 1$\,cm$^{-3}$ within the Solar circle.  The CO data cubes will be made publicly available as they are published.  


Submitted to the {\it Publications of the Astronomical Society of Australia (PASA)}, 15 May 2013.  \\
Accepted for publication  03 July 2013.

\medskip
\medskip
\end{minipage}
\end{changemargin}
]
\small

\section{Introduction}
\label{sec:introduction}
\subsection{Motivation for CO Surveys}
\label{sec:motivation}


One of the basic activities of a spiral galaxy like our own Milky Way is the continual collection of diffuse and fragmented gas and dust clouds into large Giant Molecular Clouds (GMCs), which in turn produce most of the star formation occurring within a galaxy. Surveys of the molecular component of the interstellar medium play an essential part in our understanding where and how star formation takes place in our Galaxy.  While molecular hydrogen (H$_2$) is, of course, the principal component of a molecular cloud, it is through surveys of the carbon monoxide (CO) molecule, the next most abundant molecule in the interstellar medium with several $\times 10^{-5}$ the abundance of H$_2$, that the locations of, and conditions within, molecular clouds are most readily determined.  This is because the lowest energy transition of H$_2$, the 28.2$\mu$m quadrupole J=2--0  line, arises from 510\,K above ground and so is not excited in the bulk of the molecular gas. On the other hand, the J=1--0 dipole transition of  CO, at 2.6\,mm, has an energy 5\,K above ground, so is well matched for excitation at the typical 10--20\,K temperatures found in molecular clouds.  Furthermore, the critical density at these temperatures, $\rm \sim 10^3 cm^{-3}$, is typical of the density of much of the gas \citep[see, e.g.,][for the relevant CO parameters]{2009EAS....34...89G}.

The need to use a trace species to survey the molecular medium is in contrast to surveys of the atomic medium, which can be probed directly through the 21\,cm HI line.  CO surveys generally provide much higher angular resolution than HI surveys, on account of the two order of magnitude wavelength difference resulting in smaller diffraction beam sizes.  But conversely, this makes large areal coverage much harder to obtain.  CO emission lines are relatively narrow, with line widths of a few km/s in contrast to tens of km/s for HI, so the emitting sources can be more precisely located in distance using the Galactic rotation curve.  The gas temperature can also readily be estimated for molecular gas, unlike for the HI line where it is not generally clear whether the emitting gas arises from the cold ($\sim 100$\,K) or warm ($\sim 8,000$\,K) neutral medium.

The reasons for undertaking this new CO survey of the southern Galactic plane, at higher angular and spectral resolution than previous surveys, and with greater bandpass together with multiple isotopologues, are many.  The data can be used to measure the distribution of GMC masses and sizes \citep{2009ApJ...699.1092H} in this important region of the Galaxy.  The distribution of the columns of molecular gas (derivable via the isotopologue ratios) through GMCs can be linked to the star formation rates in GMCs. Current attempts to relate molecular cloud characteristics to the star formation rate compare local molecular clouds to the Galactic Centre and external galaxies \citep[e.g.,][]{2012ApJ...745..190L, 2012ApJ...745...69K, 2013MNRAS.429..987L}; by surveying GMCs throughout the Galaxy, the size-scale between local and extragalactic may be explored. Star formation rates can then be compared to column and volume densities to test theories that predict these relationships \citep[e.g.,][]{1989ApJ...345..782M}. The turbulent velocity field can be determined.  The structure (shape and clumpiness) of GMCs can be examined \citep[e.g.,][]{2009ASPC..417...11S, 2010ApJ...723..492R}.  The star formation rates within the GMCs can be correlated with their properties.  The $\rm X_{CO}$ factor, linking CO intensity to H$_2$ column density, often the only way of estimating molecular masses, can be determined \citep[e.g.,][]{bolatto_2103}. 

The principal motivations for this survey, however, have been two-fold.  One has been to probe the connection between gamma-rays, cosmic rays and molecular gas.  The other has been to understand the formation of molecular clouds in the interstellar medium by following the evolution of elemental carbon through its ionised, atomic and molecular forms in the interstellar medium.  In \S\ref{sec:gamma} and \S\ref{sec:formationclouds} we explore these motivations further, and explain why a molecular survey with the parameters being used here is necessary in order to make further progress on these problems.

\subsection{A Brief History of \\ Southern CO Surveys}
\label{ref:history}
Galactic CO surveys have mostly been confined to strips along the Galactic plane, beginning with the pioneering surveys in the 1970's by \citet{1975ApJ...199L.105S},  \citet{1976ApJ...208..346G} and \citet{1977ApJ...217L.155C}.  On account of the relatively small beam widths of mm-wave telescopes (arcminutes) most CO mapping surveys have been targeted towards sources of particular interest, such as catalogued massive stars or infrared sources \citep[e.g., Orion by][]{1986AJ.....91.1350T}.  

The progress of unbiased surveys for the molecule's emission has been much slower and northern hemisphere surveys have preceded those of the southern hemisphere. Several surveys have now been conducted in the north \citep[e.g.,][]{1985AJ.....90..254K, 2001ApJS..136..137L, 2006ApJS..163..145J}, but progress in the south has been more limited.  Furthermore, most of the surveys have also been under sampled (with spacings between points greater than, or at best equal to, the beam size).  The first large scale survey was the Columbia survey in the first quadrant of the Galaxy using a 1.2\,m telescope in New York City \citep{1985ApJ...297..751D}.  This led, with the addition of a second telescope in Chile, to a complete Galactic Plane survey \citep{1987ApJ...322..706D}, albeit at a modest resolution of $0.5^{\circ}$.  The major structural feature of the molecular Galaxy is readily apparent -- the molecular ring, extending $\sim 60^{\circ}$ about the Galactic centre and corresponding to emission from molecular gas in the spiral arms located 3--5\,kpc from the Galactic centre.  The \citet{1987ApJ...322..706D} survey was actually a composite of 16 smaller surveys, of varying coverage and sampling, undertaken by several groups.  In particular, that covering the fourth quadrant (i.e.\ of the Galactic plane from $l = 270^{\circ} - 360^{\circ}$, located in the southern hemisphere) was undertaken by \citet{1987ApJ...314..374N}.  This survey was, in turn, improved upon by \citet{1989ApJS...71..481B}, covering the region $l = 300^{\circ} - 348^{\circ}$ and $b = \pm 2^{\circ}$, with an angular resolution of $9'$ using the same 1.2\,m telescope as Nyman et al.   The results of these early CO surveys of the Milky Way are summarised in the review article by \citet{1991ARA&A..29..195C}.

The surveys were all incorporated into an improved survey of the complete Galactic plane by \citet{2001ApJ...547..792D}.  The resulting data cubes are today the prime source of CO emission maps of the fourth quadrant available to the general astronomy community. A sparsely sampled survey (generally $4'$ spacing with a $3'$ beam) of the fourth quadrant has also been undertaken using the 4\,m Nanten telescope in Chile \citep{2008mgng.conf...11O}.

In contrast to the first quadrant, the fourth quadrant is relatively clear of low-velocity emission (associated with molecular clouds within $\sim 1$\,kpc).  Emission is generally confined to the velocity range $V_{LSR} \sim -20$ to $-100$\,km/s, and primarily associated with the Scutum--Crux spiral arm, with lesser contributions from the Norma--Cygnus and Sagittarius--Carina spiral arms \citep[see][]{2008AJ....135.1301V}.  

All of the above surveys are in the most abundant, and hence brightest, ($^{12}$CO) isotopologue of the CO molecule.  However, on account of its abundance, the $^{12}$CO line is generally optically thick, and so does not provide a direct measure of the column density of the molecule. An averaged, empirically-derived X--factor is thus used to obtain column densities from the measured CO line flux (e.g., $\rm X_{CO} = 2.7 \times 10^{20}$ $\rm \, cm^{-2} \, K^{-1} \, km^{-1} \, s$ for $|b| < 1^{\circ}$; \citet{2001ApJ...547..792D}).  The $^{13}$CO line, generally $\sim 5 - 10$ times weaker than $^{12}$CO, with a $^{13}$CO abundance $\sim 50$ times less than $^{12}$CO, provides a more accurate representation of column density when it is bright enough to be measured.  \citet{2006ApJS..163..145J} have undertaken such a survey in the $^{13}$CO line in the first quadrant (the `Galactic Ring Survey' or GRS), also achieving a much higher resolution (fully sampled at FCRAO's 14\,m telescope's $0.75'$ resolution over 75 deg$^2$).  At this resolution the structure of the individual GMCs can also be resolved.  Our survey of the fourth quadrant, with a similar angular and spectral resolution, will provide a corresponding picture of the other side of the Galactic centre, and so complement the GRS\@.

\subsection{Gamma-Rays, Cosmic-Rays \\ and Molecular Clouds}
\label{sec:gamma}
Molecular (CO) mapping data can be used to probe the connection to the ``missing'' gas inferred to exist from gamma ray (EGRET and Fermi--LAT GeV-energy maps; \citet{2005Sci...307.1292G}, \citet{2010ApJ...710..133A}) and infrared \citep{2011A&A...536A..19P} observations.  The distribution and dynamics of molecular gas also plays a pivotal role in understanding the nature of gamma-ray sources revealed by (for example) Fermi--LAT \citep{2012ApJS..199...31N} and HESS \citep{2005Sci...307.1938A, 2006ApJ...636..777A} at GeV and TeV energies respectively. This is connected to the question of the origin of Galactic cosmic-rays (CRs) since the gas acts as a target for CR collisions, leading to subsequent $>$ GeV-energy gamma-ray production. Hence measurement of these gamma-rays can provide an estimate of the total amount of gas existing in all states (i.e.\ ionized, neutral and molecular).  When used in conjunction with surveys of the atomic gas (i.e.\ HI) and of carbon in its ionized, neutral and molecular forms (i.e.\ of C$^+$, C and CO) this may allow us to infer where the ``missing'' molecular gas lies.

While there are some good examples of gamma-ray sources being associated with supernova remnants as their cosmic-rays accelerators \citep[e.g.,][]{2008A&A...481..401A, 2013Sci...339..807A}, over 30\% of Galactic gamma-ray sources remain unidentified in relation to the nature of the parent particles (i.e.\ whether they are cosmic-rays or electrons) and their counterpart accelerators \citep{2009ARA&A..47..523H, 2012ApJS..199...31N}. The moderate angular resolution of the HESS survey ($10'$--12$'$), and of the current large-scale CO surveys in the southern sky ($> 4'$) that it is compared to, have contributed to the present difficulty in identifying sources for the gamma-rays. Additionally, an arcminute scale CO survey, including line isotopologues, will provide complementary information (e.g.\ robust mass estimates) in studies of dense gas cores towards TeV gamma-ray sources \citep[e.g.,][]{2012MNRAS.419..251N}, which can be used to probe the fundamental transport properties of cosmic rays in molecular clouds \citep[e.g.,][]{2007Ap&SS.309..465G, 2012MNRAS.422.2230M}.

The southern component of the future Cherenkov Telescope Array \citep[CTA,][]{2011ExA....32..193A} will have a TeV gamma-ray sensitivity and angular resolution that are factors of  $\sim 10$ and 3 to 5 times better than that of HESS, respectively. Thus CTA will provide arcminute-scale resolution gamma-ray maps of the Galactic Plane. Moreover, CTA may detect the TeV emission from the diffuse or ``sea" of Galactic cosmic rays that are interacting with molecular clouds and/or regions of enhanced cosmic ray density in the vicinity of particle accelerators \citep{1991Ap&SS.180..305A, 2013APh....43..276A}. As such, a large-scale CO survey with arcminute resolution in the southern hemisphere will be essential for disentangling the complex gamma-ray morphologies that CTA is expected to reveal,  in a similar manner that present CO surveys (such as Dame et al.) are being applied to ($>10'$ scale) Fermi--LAT GeV data.

\subsection{The Formation of Molecular Clouds}
\label{sec:formationclouds}
Observations of external spiral galaxies show that massive stars and GMCs tend to form in the compressed regions of spiral arms, behind the spiral density wave shock.  If this region of the galaxy is primarily atomic, then the atomic gas is somehow collected together to form GMCs, as seen in the galaxy M33 \citep{2003ApJS..149..343E}.  If this region is mainly molecular, such as in our Galaxy's molecular ring, then the collection into GMCs may involve small molecular clouds (``fragments'') bound by pressure rather than self-gravity.  The manner in which gas is gathered into GMCs has yet to be determined, but four principal mechanisms have been proposed \citep{1996IAUS..169..551E}: 

\renewcommand{\labelenumi}{\roman{enumi}. }

\begin{enumerate}
\item the self gravitational collapse of an ensemble of small clouds, possibly along magnetic field lines as in the Parker instability \citep{2004ASPC..317..248O},
\item the random collisional agglomeration of small clouds \citep{1987ApJ...315...92K}, 
\item the accumulation of material within high pressure environments such as shells driven by the winds and supernovae from high mass stars \citep{1987ApJ...317..190M}, and 
\item the compression and coalescence of gas in the converging flows of a turbulent medium \citep[e.g.,][]{2000A&A...359.1124H}.  
\end{enumerate}

These scenarios provide quite different pictures for the structure and evolution of GMCs.  For instance, under mechanism (i) the gravitational collapse of a cluster of clouds produces molecular clouds that are long-lived and stable, supported against gravity by internal turbulence and magnetic fields.  This can be regarded as the classical view of a GMC \citep[e.g.,][]{1999osps.conf....3B}.  This contrasts strongly with the picture given by mechanism (iv), of compression in converging flows, where gravity plays little role.  It produces molecular clouds that are transient features \citep[e.g.,][]{2007ApJ...668.1064E}.

These four scenarios also provide different observational signatures.  If clouds form by the gravitational collapse of a cluster of small clouds (i.e.\ method i), observations should show either a roughly spherical distribution of small clouds or possibly a filamentary distribution, with the filaments following ballooned magnetic field lines out of the galactic plane (i.e.\ the Parker instability) with velocity characteristics of infall.  In addition, CO measurements can determine the mass inside any cluster radius, making it possible to compare gravitational (virial) velocities with the observed velocity dispersion of the clouds; they should be similar.  If molecular clouds form by random (no gravity) collisional coagulation of small clouds (scenario ii), the velocity field of the cluster clouds will look more random and less systematic than infall, and their velocities will exceed virial speeds.  If they are formed in wind or supernova-driven shells (iii), the shell-like morphology will be apparent.  If they formed by converging flows in a turbulent medium (iv), then we should see overall a turbulent velocity field, but local to the formation sites the velocities will be coherent (converging) and not random, and the speeds will be super-virial.

\subsection{Survey Science Requirements}
\label{sec:requirements}
Any CO survey that is undertaken represents a compromise between spatial resolution, areal coverage, spectral resolution, spectral band width and the number of isotopologues included.  Higher spatial resolution sacrifices areal coverage, however arcminute or better angular resolution is now available for infrared and millimetre continuum surveys of the Galactic plane, as conducted by e.g., Spitzer, Herschel and APEX, and so now is necessary for CO as well.  A $30''$ angular resolution would yield a spatial scale of 0.03\,pc at 200\,pc (nearby molecular clouds), 0.5\,pc at 3.5\,kpc (Galactic Ring) and 1.3\,pc at 8.5\,kpc (Galactic Centre).  The typical line width in a GMC is $\sim 5$\,km/s and in a quiescent, low mass cloud it may be as narrow as 0.1 km/s.   While the typical extent of the emission along any given sightline in the Galactic plane is $\sim 100$\,km/s, emission is detected over a $\pm 250$\,km/s range through the Galaxy (including, in particular, over this full velocity range for fields within a few degrees of the Galactic centre).  

\paragraph{Areal Coverage and Spatial Resolution}
~ \\
\noindent
The assembly time for a GMC is approximately the radius ($\gtrsim 100$\, pc) of a cluster of small clouds divided by the speed ($\sim 5$\,km/s, turbulent or gravitational) at which they come together; i.e.\ $\gtrsim 20$\,Myr.  Since this is comparable to, or greater than, the estimated ages of GMCs, it is necessary to observe as many, or more, clusters as there are GMCs along each line of sight (i.e.\ $\sim 1 - 10$).  A suitable survey area should include at least 100 GMCs and so include at least this number of ``forming'' GMCs.  

\begin{figure*}[hp]
\begin{center}
\includegraphics[scale=0.6]{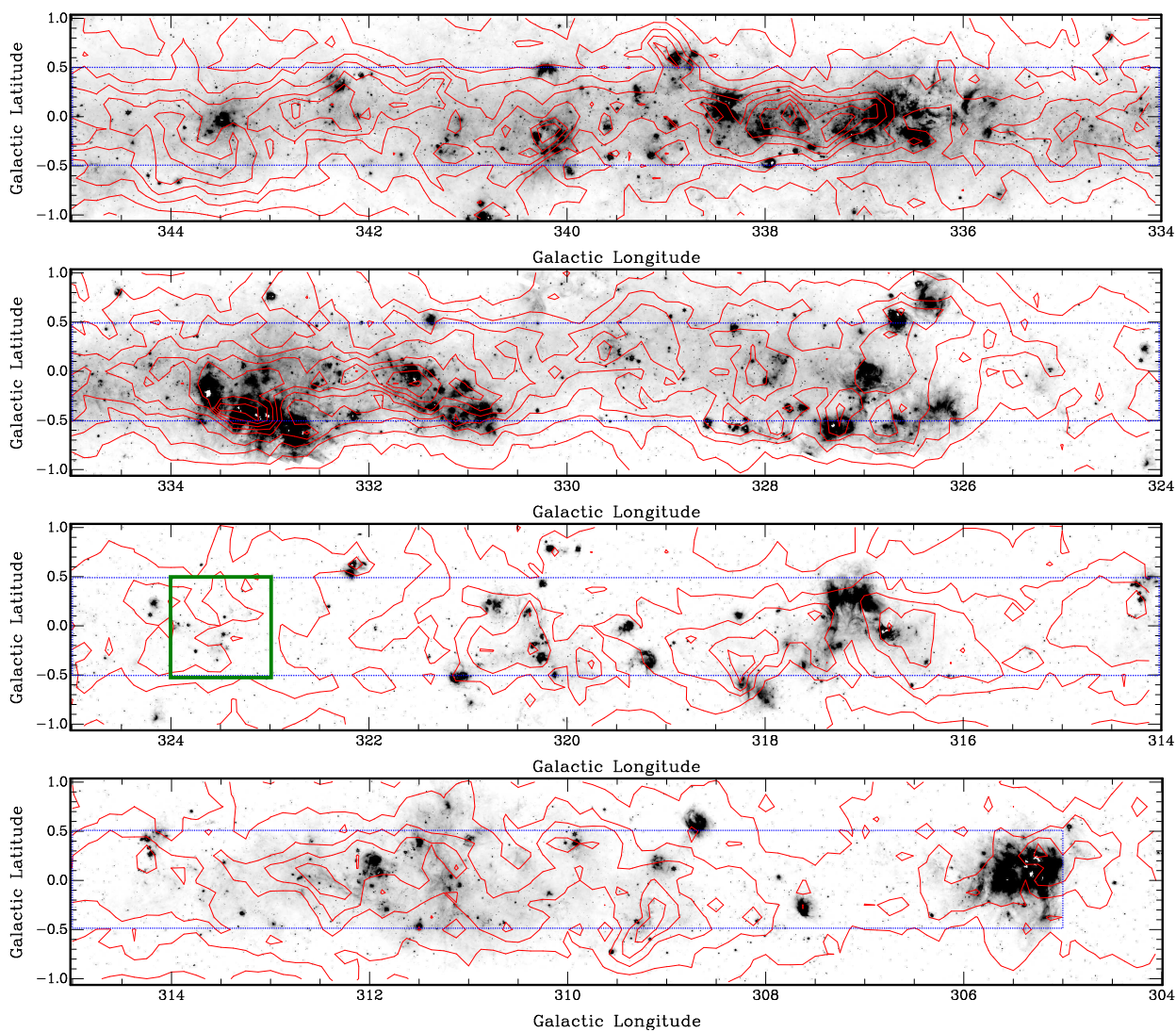}
\caption{Spitzer/MIPSGAL 24$\mu$m image of the Galactic plane \citep{2009PASP..121...76C}, shown as a series of $11^{\circ} \times 2^{\circ}$ panels (with $1^{\circ}$ overlap between each), overlaid with red contours of $^{12}$CO emission from the \citet{2001ApJ...547..792D} survey. The region planned for our Mopra survey, from $l = 305^{\circ}$ to  $345^{\circ}, b = \pm \,0.5^{\circ}$, is indicated with the blue dotted lines.  The data included in this paper, from the G323 region, comes from the region indicated by the solid green box.}
\label{fig:survey}
\end{center}
\end{figure*}

In order to contain sufficient gas to build a molecular cloud, an interstellar cloud needs to have a hydrogen column of order $\rm 10^{21} \, cm^{-2}$ in order to shield any molecules that form within it from dissociating UV radiation.  This corresponds to a diameter of about 7~pc at interstellar pressures \citep{2003ApJ...587..278W}.  Small molecular clouds also need similar columns, but are cooler and denser than atomic clouds and so may have sizes of order 1 to 2\,pc.  GMCs themselves have diameters of 10 to 100\,pc.  Ensembles of small clouds that are moving at detectable (i.e.\ $> 1$\,km/s) speeds to coalesce into GMCs will have ensemble diameters of 200 to 1000\,pc.  The survey area (see Fig.~\ref{fig:survey}) goes through the molecular ring of our Galaxy at distances of typically 4 to 8\,kpc, and therefore these sizes correspond to $0.5' - 2'$ for the small clouds, $4' - 80'$ for the GMCs, and $1^{\circ} - 10^{\circ}$ for the ensembles of small clouds.  Therefore, it is necessary to have at least $0.5'$ spatial resolution to resolve individual clouds and a survey area that extends across a spiral arm and is at least three times larger than the ensemble size (i.e.\ $\sim 30^{\circ}$) to ensure that it encompasses the complete range of phenomena that are occurring. 

\paragraph{Spectral Resolution and Bandwidth}
~ \\
\noindent
Line widths observed toward small individual clouds are of order 1 km/s and toward GMCs $\sim 5$\,km/s.  Velocity information is generally used to place the clouds along the line of sight, using the galactic rotation curve.  In the $l = 323^{\circ}$ direction 1 km/s corresponds to about 90\,pc.  However, if clusters of clouds are seen in the two dimensions on the sky, then it is possible to also determine their velocity dispersion by eliminating any spatial elongation along the sight lines \citep[this is akin to the ``finger of God'' structures seen in galaxy redshift surveys; e.g.,][]{1994ApJ...422...46P}.  Along several sight lines observed using Mopra by the GOT C+ Herschel program  5 to 10 CO features are typically seen \citep{2010A&A...521L..17L}.  Therefore, it is necessary to have $< 1$\,km/s spectral resolution to determine the 3D distribution of clouds and 0.1 km/s resolution to enable the velocity distributions of clouds within ensembles. Furthermore, a bandpass of 500 km/s is needed to ensure measurement over the complete range of $V_{LSR}$ velocities encountered in the Galaxy. 

\paragraph{Number of Isotopologues}
~ \\
\noindent
Up to three CO isotopologues can generally be detected in molecular clouds ($^{12}$CO, $^{13}$CO and C$^{18}$O; the very weak C$^{17}$O line requires long integrations).  The $^{12}$CO line provides the greatest sensitivity, able to reach to the threshold for CO to exist in diffuse gas, where the CO column density may be $\rm \sim 10^{14} \, cm^{-2}$\citep{1986ApJS...62..109V}.  $^{13}$CO best reproduces the column density, the line generally being optically thin.  Detection of the C$^{18}$O line allows validation of any assumptions made about the $^{13}$CO opacity. Measuring all these isotopologues is thus valuable.  To do so simultaneously requires a spectrometer whose bandwidth is at least 6 GHz.

\subsection{Synopsis of this Paper: \\ the Mopra CO Survey}
\label{sce:introsummary}
In this paper we report the first observations from a new CO survey designed to cover much of the fourth quadrant of the Galaxy which achieves the specifications discussed in \S\ref{sec:requirements}.  The survey uses the 22\,m Mopra telescope in Australia, and with an 8\,GHz  wide bandpass correlator and `on-the-fly' mapping is able to measure the four CO isotopologues with over 400 km/s bandwidth on each, 0.1 km/s spectral resolution, fully-sampled at $35''$ spatial resolution.  We present here the data from the first degree of the survey, $l = 323^{\circ} - 324^{\circ}$, $b = \pm 0.5^{\circ}$, and describe the specifications and characteristics of the full survey.  The survey is ongoing, and is planned to cover $l = 305^{\circ} - 345^{\circ}$ (see Figure \ref{fig:survey}, which overlays the \citet{2001ApJ...547..792D} $^{12}$CO integrated flux contours on the Spitzer MIPSGAL 24$\mu$m image of the southern galactic plane\citep{2009PASP..121...76C}).  An additional region around the Central Molecular Zone ($l = 358^{\circ} - 003^{\circ}$) is also being mapped.  These data sets will be made publicly available through the CSIRO ATNF data archive\footnote{They will also be accessible via this project's web page at URL http://www.phys.unsw.edu.au/mopraco.}.  The present survey provides a companion data set to that obtained by the HOPS program of the fourth quadrant, also conducted with Mopra by \citet{2011MNRAS.416.1764W}, of the 20--28 GHz band lines at $2.5'$ resolution (mainly of NH$_3$ thermal and H$_2$O maser lines).

In the following sections, \S\ref{sec:observations} describes the observations undertaken and \S\ref{sec:datareduction} the data reduction techniques employed, \S\ref{sec:dataquality} discusses the quality of the data, \S\ref{sec:interpretingco} discusses how CO data may be interpreted in terms of physical parameters for the emitting sources, \S\ref{sec:results} presents the results from our analysis and \S\ref{sec:discussion} interprets them.  Finally, \S\ref{sec:conclusions} summarises the principal conclusions of this work.

\section{Observations}
\label{sec:observations}
The observations we discuss here, maps along the southern Galactic plane of the four isotopologues of the CO J=1--0 line ($^{12}$C$^{16}$O, $^{13}$C$^{16}$O, $^{12}$C$^{18}$O and $^{12}$C$^{17}$O; hereafter simply $^{12}$CO, $^{13}$CO, C$^{18}$O and C$^{17}$O),  were conducted using the 22\,m diameter Mopra millimetre-wave telescope of the CSIRO Australia Telescope National Facility, sited near Coonabarabran in NSW, Australia.  We used the UNSW Mopra Spectrometer (MOPS) digital filterbank and the 3\,mm band receiver. The MMIC (Monolithic Microwave Integrated Circuit) receiver covers the spectral range from 77 to 117\,GHz and the 8\,GHz bandpass of the MOPS was centred at 112.5\,GHz to include these four isotopologues.  The angular resolution of the Mopra beam size is $33''$ FWHM \citep{2005PASA...22...62L}, and after the median filter convolution applied in the data reduction is around $35''$ in the final data set.  The extended beam efficiency, $\eta_{XB} = 0.55$ at 115\,GHz, was also determined by Ladd et al.  This is used to convert brightness temperatures into line fluxes for determination of source parameters (rather than the main beam efficiency factor, of $\eta_{MB} = 0.42$).

The data set presented in this paper was obtained in March 2011 and used the MOPS in its ``zoom'' mode of operation (in contrast to its 8\,GHz wide ``broad band'' mode), with $4 \times 137.5$\,MHz wide, dual-polarization bands, each of 4,096 channels, yielding a spectral resolution of $\rm \sim 0.09\, km\,s^{-1}$.  Table~\ref{tab:lineparams} presents the parameters used for the line measurements.  This survey is ongoing; data has also been collected over the Austral winters of 2011 \& 2012 and this is planned to continue in subsequent years.  In 2012 the number of zoom modes was increased from 4 to 8 and for completeness we also include the corresponding line parameters in Table~\ref{tab:lineparams}.

The observations were conducted using `fast on-the-fly' mapping (FOTF), which is a modification of the standard OTF mapping procedure in order to allow larger areas to be mapped (for a corresponding reduction in integration time per position).  Table~\ref{tab:scanparams} summarises the scan parameters.  The telescope is scanned in one direction (either Galactic $l$ or $b$) for $1^{\circ}$ at a rate of $\rm 35''/s$.  Each 2.048\,s cycle time of the system is divided into 8 bins of 256\,ms in which the data acquired in that interval is recorded (known as the `pulsar binning mode').  This yields a cellsize of $9''$ in 1 bin, approximately one-quarter of the beam size.  50 cycles are required to cover $1^{\circ}$, taking approximately 2 minutes of clock time.  A sky reference position is then observed for 7 cycle times ($n \sim \sqrt50 $ so equivalent to 14.3\,s of integration in order to optimally reduce the noise contribution from the reference position), for later subtraction from each cell position along the scan.  The telescope is then scanned in the opposite direction with a row/column spacing of $15''$, followed by another reference position measurement.  This procedure is repeated 24 times until a region of $60' \times 6'$ has been mapped (1 ``footprint").   A paddle calibration measurement (ambient temperature load) is made every 25 minutes, to place the data on the $T_A^*$ (K) scale.  In total, with telescope overheads, approximately 1 hour is needed to complete this footprint.  A bright SiO 86\,GHz maser (AH Sco, W Hya or VX Sgr) is then observed to determine the pointing corrections for the next footprint.  These offsets were typically found to be between $5''$ and $10''$.  

Each night a standard reference source (M17 SW; RA=18:20:23.2, Dec=$-16$:13:56 J2000) was also observed with the CO line spectral configuration in order to monitor the overall system performance.  While this source was observed over a range of air masses, the peak brightness found, $T_A^* = 29$\,K, is equivalent to $T_{MB} = 53\,$K when correcting for the extended beam efficiency.  By comparison,  the peak brightness temperature for M17 SW measured by the SEST telescope, when this source was a part of its calibration monitoring program, was $T_A^* = 40.4$\,K.  This is equivalent to $T_{MB} = 51\,$K after correcting for an estimated extended beam efficiency of $\eta_{XB} = 0.8$.\footnote{A main beam efficiency for SEST of $\eta_{MB} = 0.7$ is recorded on the archived web pages at URL http://www.apex-telescope.org/sest/, from which we estimate $\eta_{XB} \sim 0.8$.}  The peak emission region for M17 SW extended over $\sim 2'$, so that the different beam sizes of Mopra and SEST ($35''$ c.f.\ $45''$) is not significant here.  Thus the flux scale for Mopra is similar to that of SEST\@.

For this survey the Galactic plane is mapped in 1 degree survey blocks of longitude ($l$), each extending $\pm 0.5^{\circ}$ in latitude ($b$).  10 footprints are required scanning in the $l$-direction, and a further 10 scanning in the $b$-direction, to map this area.  In typical weather conditions this requires 4 transits of the source (`4 nights') to accomplish.  The data set presented here is from the G323 block (i.e.\ $l=323-324^{\circ}, b=\pm 0.5^{\circ}$), and was the first region to be mapped in this survey of the southern Galactic plane.  A sky reference position, chosen to be free of CO emission, at $l=323.5^{\circ}, b=-2.0^{\circ}$ was used.  Our intention is to map the region from $l=305^{\circ}-345^{\circ}$.  At the time of writing data for $l=323^{\circ} - 340^{\circ}$ has been obtained.  A separate program is also being undertaken to map the extensive CO emission from the Central Molecular Zone, covering the region $l=358-003^{\circ}, b=-0.5$ to $+1^{\circ}$; this will be reported upon separately\footnote{See also http://www.phys.unsw.edu.au/mopracmz and \citet{2012MNRAS.419.2961J}, \citet{2012MNRAS}.}.  

\section{Data Reduction}
\label{sec:datareduction}

There are four stages to the data reduction.  First, each spectrum needs to be tagged with its angular position and combined with the nearest reference position measurement.  Secondly, the data are interpolated onto a uniform angular grid, taking into account overlapping beam positions.  Thirdly, the data are cleaned, both for bad pixels and for poor rows or columns (for instance, caused by poor weather during the reference measurement).  Finally, the data are continuum subtracted, to yield data cubes for each spectral line, of the brightness temperature as a function of galactic coordinate and $V_{LSR}$ velocity.  The first two steps use the {\sc livedata} and {\sc gridzilla}\footnote{See http://www.atnf.csiro.au/computing/software/livedata/} packages developed by Mark Calabratta at the CSIRO--ATNF\@.  The latter two use custom-written IDL\footnote{See http://www.exelisvis.com/ProductsServices/IDL.aspx.} routines.

\begin{figure*}[h]
\includegraphics[scale=0.45, angle=0]{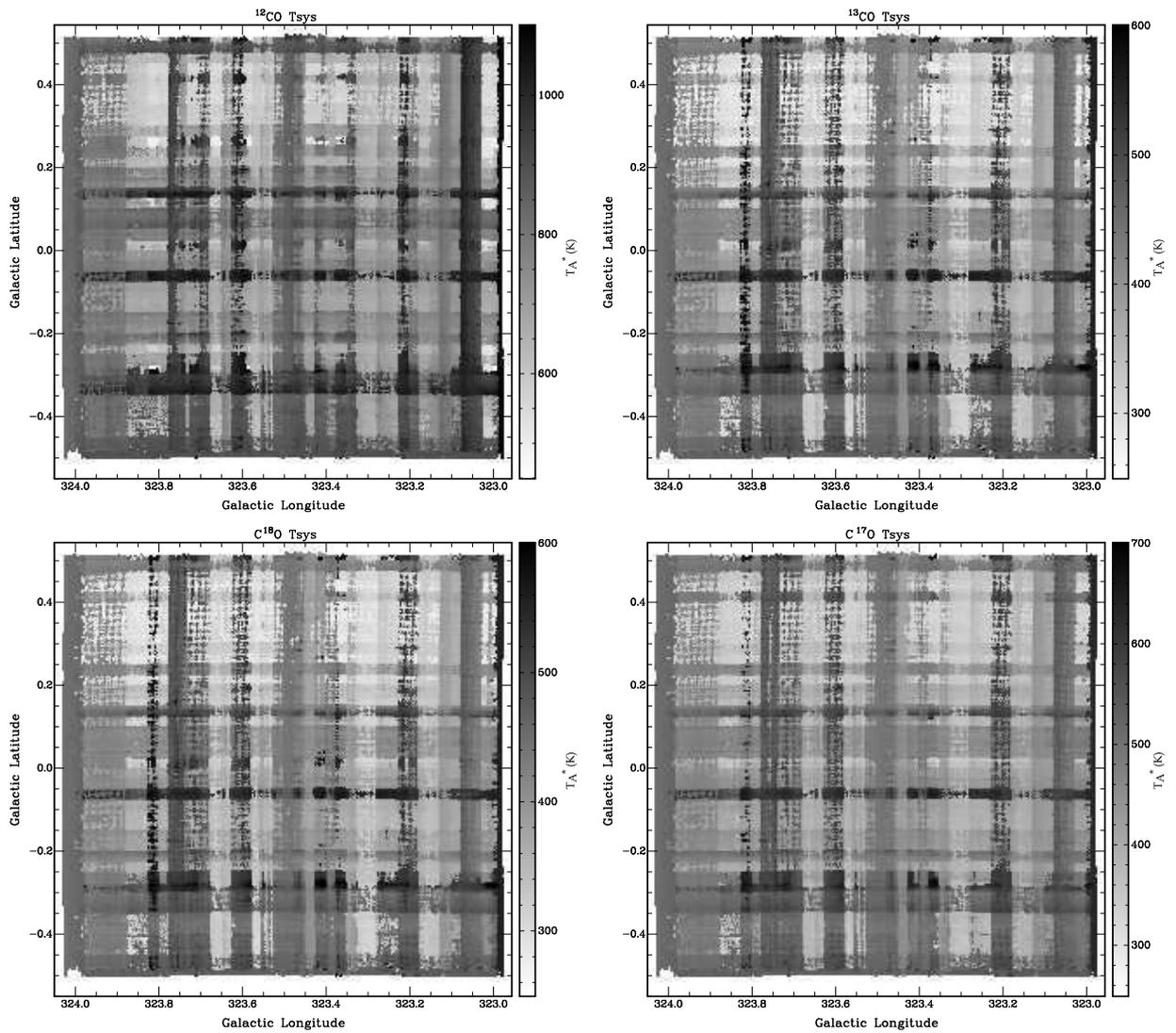}
\caption{$T_{sys}$ images for, from top-left going clockwise: $\rm ^{12}CO, ^{13}CO, C^{17}O \, \& \, C^{18}O$, in units of K (as indicated by the scale bar).  The striping pattern is inherent in the data set and results from averaging the scanning in the $l$ and $b$ directions in variable weather conditions.   }
\label{fig:tsys}
\end{figure*}

{\sc livedata} takes the raw data in {\sc rpfits}\footnote{See http://www.atnf.csiro.au/computing/software/rpfits.html.} format, bandpass corrects and calibrates them using the nearest reference spectrum, subtracts a linear baseline (masking out 400 channels on either edge of the bandpass before calculating this), with the output formatted as {\sc sdfits} \citep{2000ASPC..216..243G} spectra.  Using the {\sc gridzilla} program, these are then gridded into data cubes with a $15''$ grid spacing and $262 \times 262$ spatial positions, over the velocity extent covered for each line and centred on its rest $V_{LSR}$ velocity.  A median filter is used for the interpolation of the over-sampled data onto each grid location, as this is more robust to outliers than the averaging option available in {\sc gridzilla}.  Spectra for which the $T_{sys}$ value lies outside the range [400\,K, 1,000\,K] for the $^{12}$CO cube and [200\,K, 700\,K] for the other lines are excluded during this process (see \S\ref{sec:dataquality} and Fig.~\ref{fig:tsys}). The output is a {\sc fits} format cube in ($l, b, V_{LSR}$) coordinates.

Cleaning the data cubes involves two steps,  firstly identifying individual bad pixels and secondly bad rows or columns.  In both cases the relevant pixels are replaced by interpolating the values of neighbouring pixels, carried out using purpose-written IDL routines.  Isolated bad pixels are identified as being more than $5 \sigma$ different than the mean of a $5 \times 5 \times 5$ box centred on them. The cleaning process is iterated until no more bad pixels are found; the number found amounted to $\sim 0.01\%$ of the total. 

Bad rows or columns are identified by determining the median value of each row (or column), when summed over a velocity range representative of the continuum, and comparing this to the median of the entire data set (over the same velocity range).  If it differs by more than $3 \times$ the standard deviation of the entire data set then that row (column) is interpolated over using an inverse distance weighting algorithm. This uses the values of the ``good'' pixels within a ring of radius 2 pixels around those of the bad rows and columns, weighting the distribution so that the pixels closest have greater influence over the interpolated point (using a power parameter $p = 3$).  The entire process is iterated until no further changes occur; in practice no more than 7 iterations were found to be needed, with the total number of bad rows and columns replaced being [37, 3, 7, 25] for [$\rm ^{12}CO, ^{13}CO, C^{18}O, C^{17}O$] in the G323 block, respectively.

The data cubes are then binned in {\sc miriad}\footnote{See http://www.atnf.csiro.au/computing/software/miriad.} to a $30''$ grid spacing, over $131 \times 131$ spatial positions.  This also makes them a manageable size for further analysis. The data cubes are then continuum subtracted using a custom-written IDL routine that fits a fourth-order polynomial to selected velocity ranges for each line profile, and then a seventh-order polynomial to the continuum subtracted line profiles at all data points that remain `close' (typically within 0.5 $\sigma$) to zero. 

\section{Data Quality}
\label{sec:dataquality}

In this section we provide several figures of merit for assessing the quality of the data obtained.

\begin{figure*}
\begin{center}
\includegraphics[scale=0.9, angle=0]{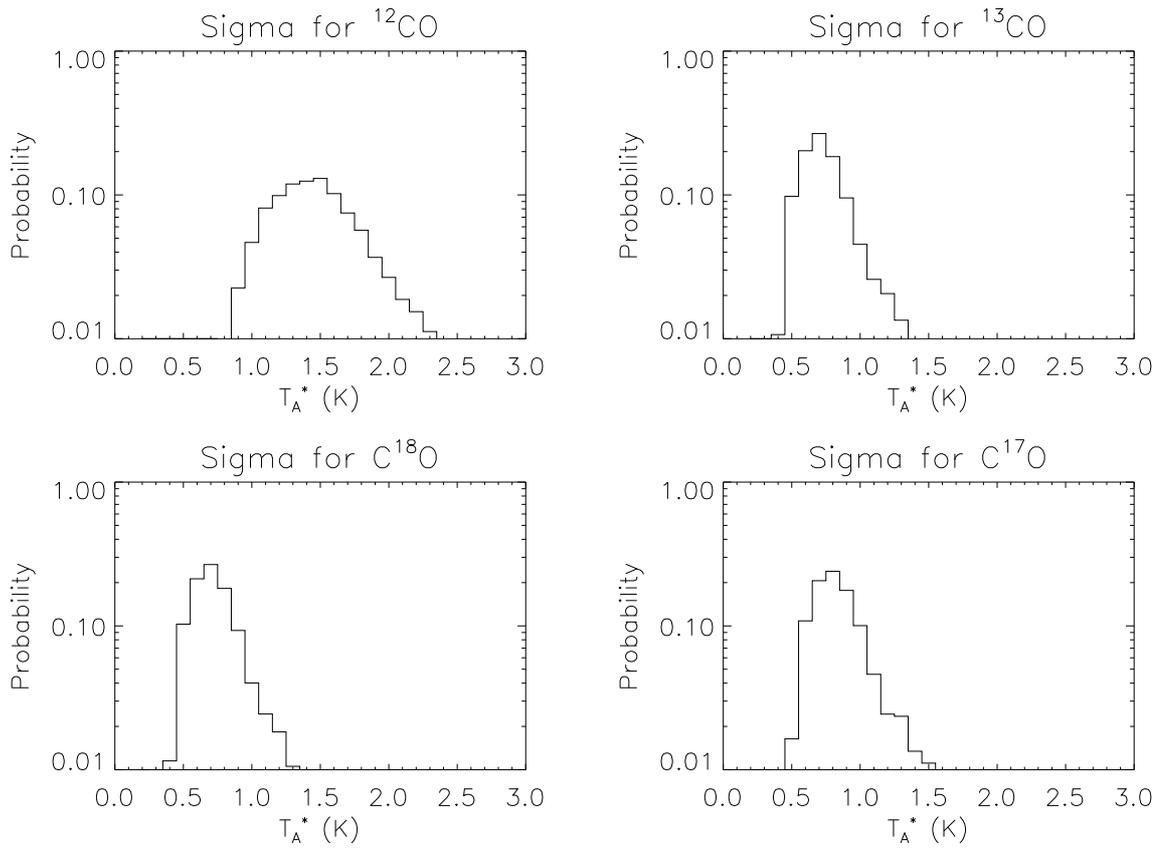}
\caption{Probability distribution of the noise level, $\sigma_{\rm cont}$, as determined from the standard deviation in the continuum channels (in $T_A^*$ (K) units) for each pixel.  From top-left, going clockwise: $\rm ^{12}CO, ^{13}CO, C^{17}O \, \& \, C^{18}O$\@.}
\label{fig:sigma}
\end{center}
\end{figure*}

The noise per channel, $\sigma_{\rm cont}$, is determined from the standard deviation of the channels outside the range where line emission occurs.  This was selected as the channels with velocities $\rm < -115 \, km\,s^{-1}$ and $\rm > +20 \,km\,s^{-1}$ $V_{LSR}$\@.  Histograms showing the probability distributions for the four CO lines are shown in Fig.~\ref{fig:sigma} with standard deviations and mode values for the continuum channels listed in Table~\ref{tab:lineparams}. Mode values (the $1 \sigma$ sensitivity) for the $\rm ^{12}CO$ and $\rm ^{13}CO$ lines are 1.5\,K and 0.7\,K per 0.1\,km/s velocity channel, respectively. At 115\,GHz the atmosphere is both considerably worse than at 112\,GHz (being on the edge of a molecular oxygen absorption line) and the surface of the dish is at its limit of useability.  This is evident in the noise values for $\rm ^{12}CO$ being a factor of two higher than for the other three lines. The tail to the distribution at higher noise levels results from observations in poorer conditions.

\begin{figure*}
\begin{center}
\includegraphics[scale=0.45, angle=0]{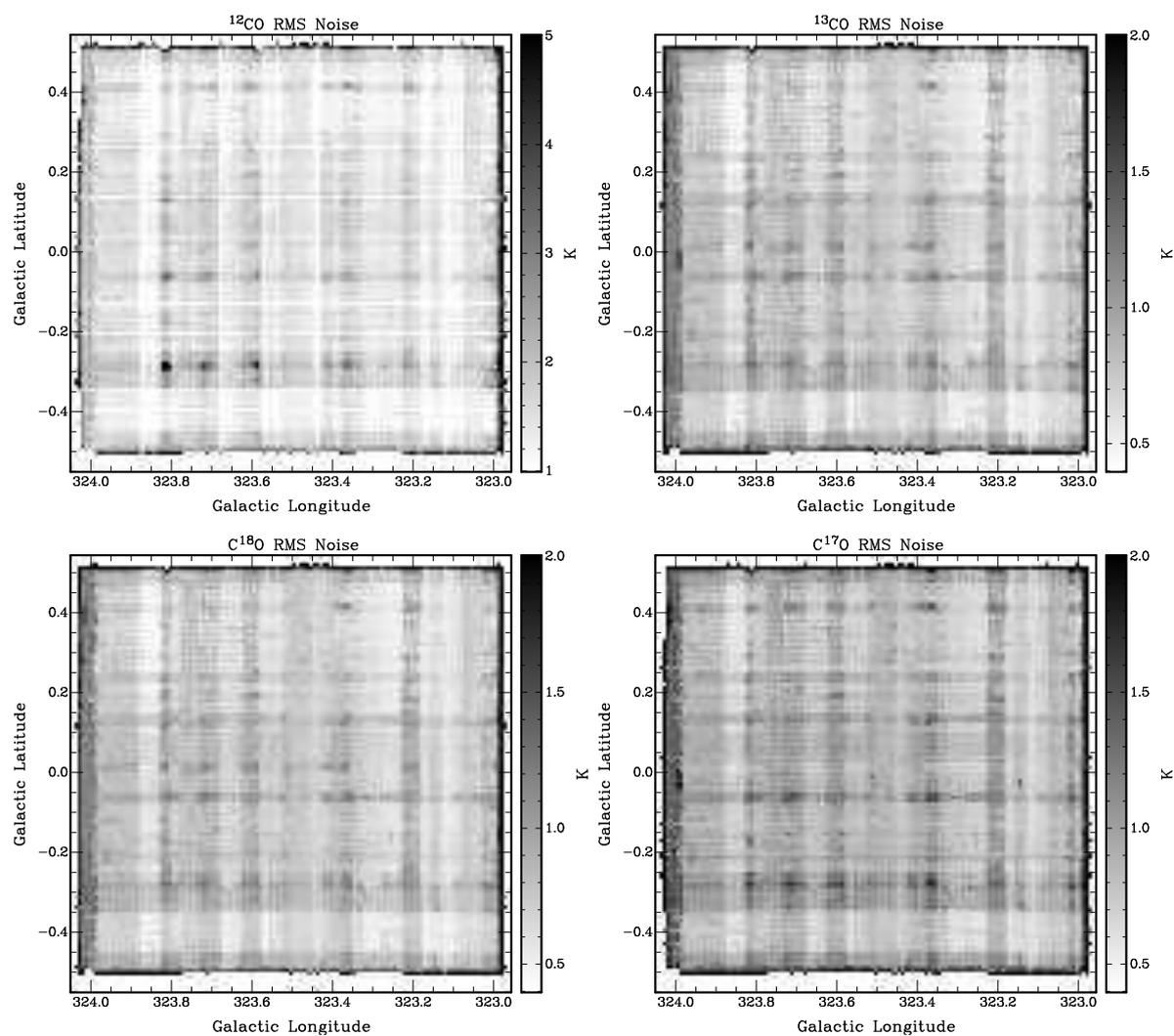}
\caption{Images showing the noise level (in $T_A^*$ (K) units) for each spectral line, determined from the standard deviation of the continuum channels between $0$ and $\rm +90\,km s^{-1}$for each pixel.  From top-left, going clockwise: $\rm ^{12}CO, ^{13}CO, C^{17}O \, \& \, C^{18}O$\@.  }
\label{fig:sigmaimage}
\end{center}
\end{figure*}

We also show in Figure \ref{fig:sigmaimage} the noise images for each of the four spectral line data cubes.  These were determined from the standard deviation, for each pixel, of the continuum in each profile (i.e.\ outside the range of the $^{12}$CO line emission; chosen to be from $0$ to $\rm +90\,km s^{-1}$). These images represent the $1 \sigma$ rms noise achieved per pixel.  They result from the combination of the data taken with a variety of system temperatures during each of the several scans obtained across that pixel position, in both the $l$ and $b$ directions (see below), as well as from the pixels, rows and columns identified as `bad' and interpolated over (as discussed in \S\ref{sec:datareduction}).   The later show up as lines with lower apparent noise because they result from the averaging of the data over several nearby pixels. As with the histograms shown in Fig.~\ref{fig:sigma}, the poorer performance for $^{12}$CO is apparent.

\begin{figure*}[h]
\begin{center}
\includegraphics[scale=0.9, angle=0]{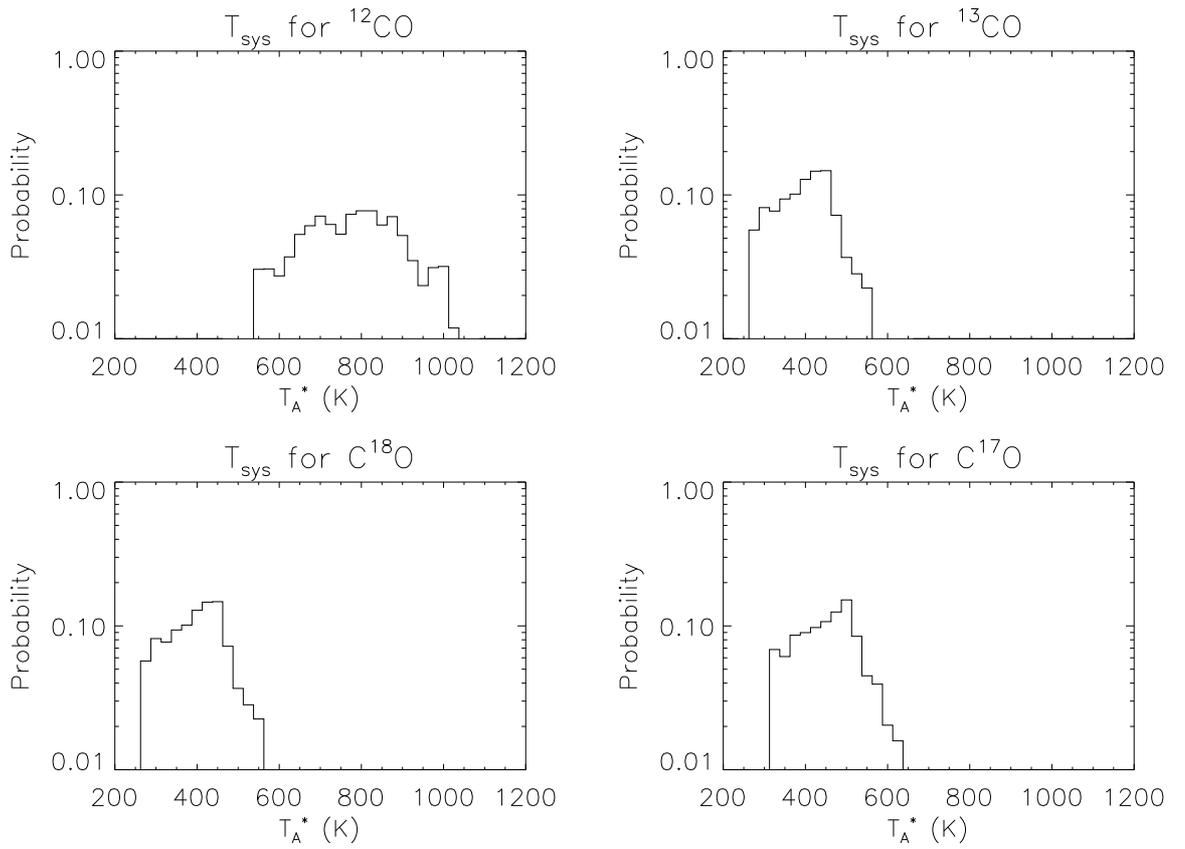}
\caption{Probability distribution of the system temperature, $T_{sys}$ (in $T_A^*$ (K) units) in the data for each pixel, determined from the ambient temperature load paddle measurements.  From top-left, clockwise: $\rm ^{12}CO, ^{13}CO, C^{17}O \, \& \, C^{18}O$\@.}
\label{fig:tsyshist}
\end{center}
\end{figure*}

The system temperature, $T_{sys}$, measures the level of the received signal, with contributions from source, sky, telescope and instrument.  It is determined through calibration with the ambient temperature paddle, which is periodically (every $\sim 30$ minutes) placed in front of the beam to cover it.  Histograms showing its probability distribution are shown in Fig.~\ref{fig:tsyshist}, with median and mode values listed in Table~\ref{tab:lineparams}.  For $\rm ^{12}CO$ and $\rm ^{13}CO$ the median values are 800\,K and 420\,K, respectively.  The inferior conditions at 115\,GHz result in $T_{sys}$ for the $\rm ^{12}CO$ line also being about a factor two higher than for the other three lines.

Images showing how $T_{sys}$ varies between pixels for the four CO lines are shown in Fig.~\ref{fig:tsys}.  The crossed striping pattern that is evident results from averaging the values from the two orthogonal scan directions and the inherent variability of the sky emission, especially in the summer period when the G323 data set was obtained.  By scanning in two directions artefacts arising from poor sky conditions are minimised.  Note also that data with excessive $T_{sys}$ is thresholded out prior to gridding (see \S\ref{sec:datareduction}), and that particularly poor footprints were repeated (and thus do not appear in the data set).

\begin{figure*}[h]
\includegraphics[scale=0.45, angle=0]{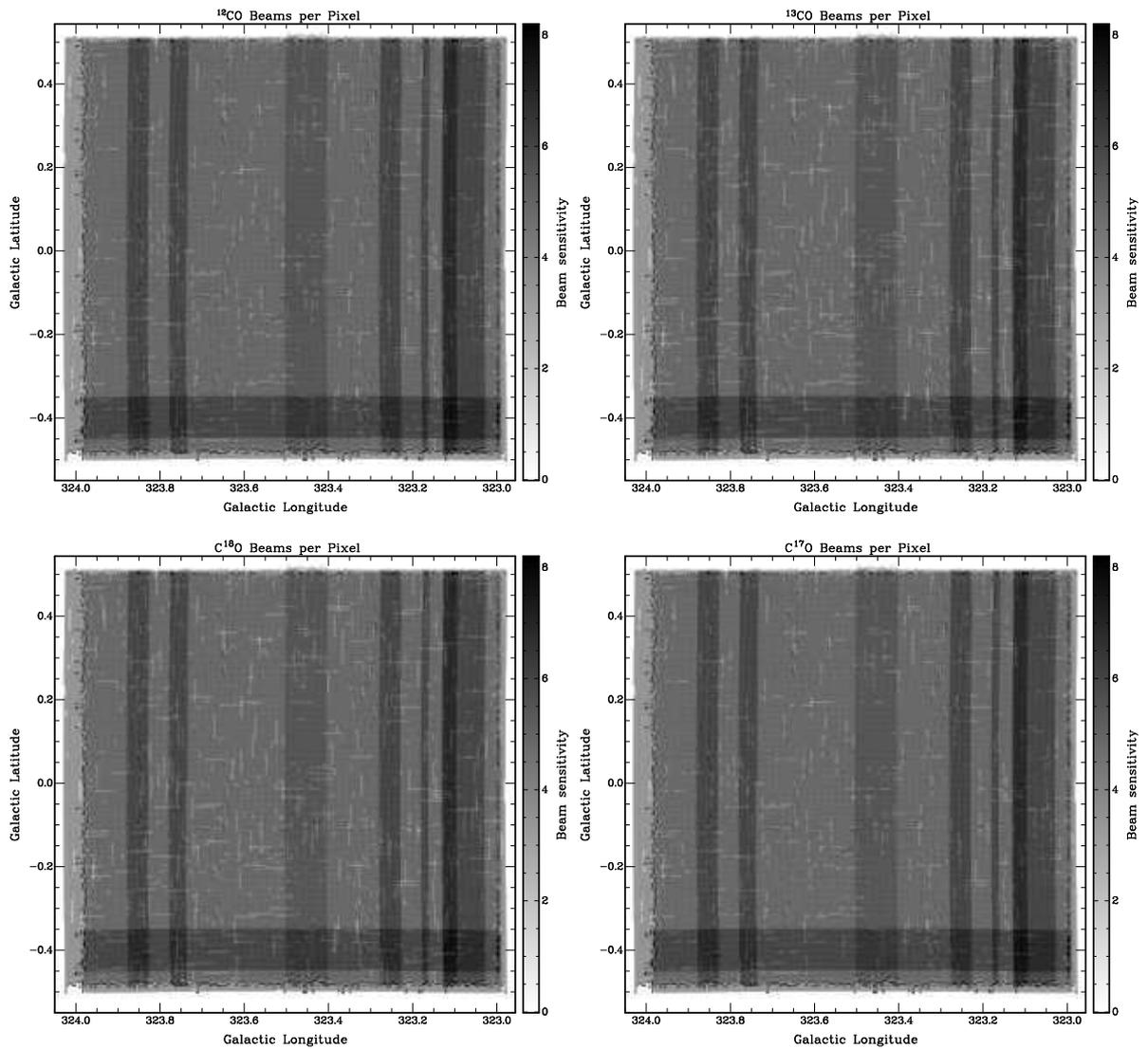}
\caption{Beam coverage images, from top-left going clockwise: $\rm ^{12}CO, ^{13}CO, C^{17}O \, \& \, C^{18}O$. These show the effective number of measurements (i.e.\ cells) that are combined per pixel position, as indicated by the scale bar.}
\label{fig:beams}
\end{figure*}

The beam coverage is shown in Fig.~\ref{fig:beams}.  This is the effective number of beams (each resulting from a single cell; see Table~\ref{tab:scanparams}) that have been combined to yield each pixel value in the final data cube. This generally is $\sim 5$ cells, but varies from line to line and region to region as a result of both thresholding of poor data (in particular for $\rm ^{12}CO$) and the number of additional footprints observed.

\section{Interpreting CO Line Data}
\label{sec:interpretingco}

\subsection{CO Brightness Temperatures and Line Ratios}
\label{sec:calcs}
As an aid to interpreting the measured CO brightness temperatures and isotopologue line ratios that are shown in the Results section (\S\ref{sec:results}), we here provide a number of figures that show calculations of these quantities and their values in relation to the optical depth in the $\rm ^{12}CO$ line.

Following the description given in \S3.7 of \citet{2012MNRAS.419.2961J}, the ratio $R_{12/13}$ of the brightness temperatures of the $\rm ^{12}CO$ and  $\rm ^{13}CO$ lines is given by
\begin{equation}
R_{12/13} = \frac{T_A^*(^{12}\rm CO)}{T_A^*(^{13}\rm CO)} = \frac{1 - e^{-\tau_{12}}}{1 - e^{-\tau_{13}}},
\label{eqn:opticaldepth}
\end{equation}
where $\tau_{12}$ and $\tau_{13}$ are the optical depths of the $\rm ^{12}CO$ and $\rm ^{13}CO$ lines, respectively (it is assumed that the lines arise from the same gas, with an assumed constant excitation temperature, $T_{ex}$).  Furthermore, if the isotope abundance ratio $X_{12/13} = [^{12}{\rm C} / ^{13}{\rm C}]$ (assumed to be equal to the isotopologue ratio for the corresponding CO lines), then $\tau_{13} = \tau_{12} / X_{12/13}$.  In the limit when the $\rm ^{12}CO$ line is optically thick and the $\rm ^{13}CO$ line optically thin (i.e.\ $\tau_{12} > 1$ and $\tau_{13} < 1$) then we obtain
\begin{equation}
R_{12/13} \sim \frac{X_{12/13}}{\tau_{12}}
\label{eqn:opticaldepthapprox}.
\end{equation} 
This is the normal situation for most data measured for these lines from molecular clouds.  

\begin{figure*}[h]
\begin{center}
\includegraphics[scale=1.0, angle=0]{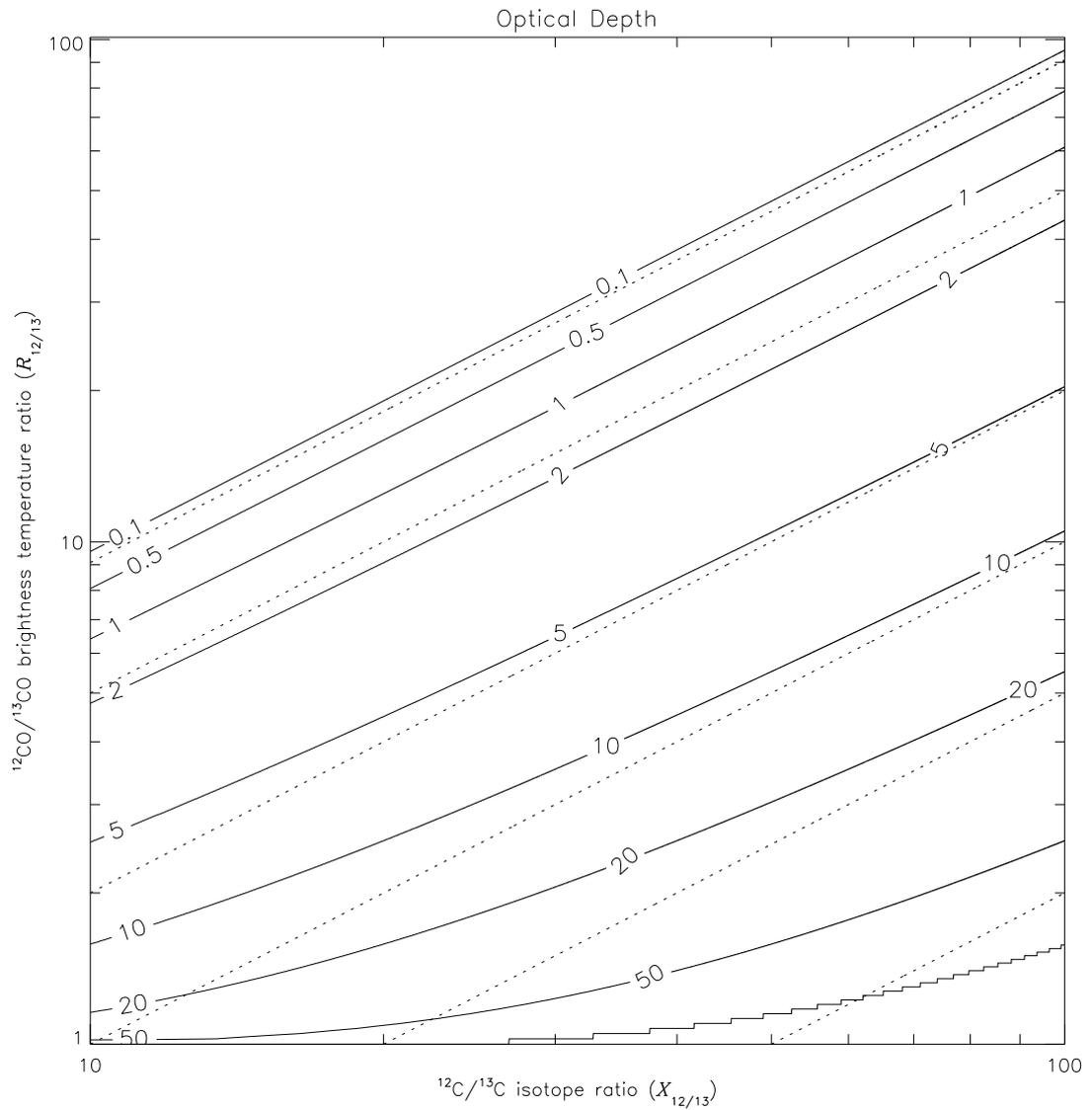}
\caption{Contour plot showing the optical depth the $^{12}$CO line as a function of the $\rm ^{12}C / ^{13}C$ isotope ratio ($X_{12/13}$) and the $\rm ^{12}CO / ^{13}CO$ brightness temperature ratio ($R_{12/13}$).  The solid lines show the full solution to equation~\ref{eqn:opticaldepth}, with contour levels (from top-left to bottom-right) drawn at $\tau = $ 0.1, 0.5, 1, 2, 5, 10, 20, 50 \& 100 (as labelled).  Dotted lines are for the limit when the $\rm ^{12}CO$ line is optically thick and the $\rm ^{13}CO$ line optically thin (see equation~\ref{eqn:opticaldepthapprox}), and drawn (but not labelled) at $\tau = $1, 2, 5, 10, 20 \& 50.}
\label{fig:opticaldepth}
\end{center}
\end{figure*}

Figure~\ref{fig:opticaldepth} presents contour plots showing the optical depth, $\tau_{12}$, as a function of the isotope ratio, $X_{12/13}$, and the brightness temperature ratio, $R_{12/13}$, with the full solution to equation~\ref{eqn:opticaldepth} shown as the solid lines and the thick/thin approximation (eqn.~\ref{eqn:opticaldepthapprox}) overlaid as dotted lines.  Note that in the limit where both lines are optically thin then $R_{12/13} = X_{12/13}$ for all $\tau < 1$. This is reflected by the linear relation for the $\tau = 0.1$ line in the plot.  Similarly, when both lines are optically thick then $R_{12/13} = 1$ for $\tau >> 1$ and this is reflected in the horizontal lines at $R_{12/13} \sim 1$ for $\tau = 50$ (when $X_{12/13}$ is low) in the Figure.  For most of the parameter space, when $\tau \gtrsim 2$, the optically thick / thin approximation yields a good solution, as seen by the dotted lines then closely following the dashed lines.

\begin{figure*}[h]
\begin{center}
\includegraphics[scale=1.0, angle=0]{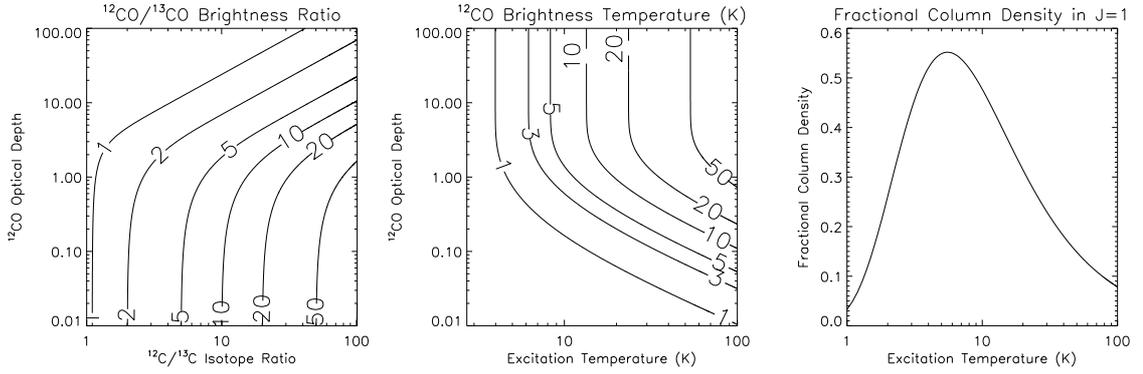}
\caption{Left panel: contour plot of the $\rm ^{12}CO / ^{13}CO$ brightness temperature ratio as a function of $\rm ^{12}C / ^{13}C$ isotope ratio and $\rm ^{12}CO$ optical depth, as derived from eqn.~\ref{eqn:opticaldepth}.  Contour lines are labelled and are for ratios of 1, 2, 3, 10, 20 \& 50, respectively. 
Middle panel: contour plot of the $\rm ^{12}CO$ brightness temperature (in K) as a function of the excitation temperature (in K) and the $\rm ^{12}CO$ optical depth, as derived from eqn~\ref{eqn:temp}.  The contour lines are labelled as follows: $T = $1, 3, 5, 10, 20 \& 50\,K\@. 
Right panel: graph showing the fractional column density in the $J=1$ level of the CO molecule as a function of the excitation temperature (in K), as given by the Boltzmann equation (eqn.~\ref{eqn:fraccolumn}).  The peak occurs for $T_{ex} = 5.5$\,K, the energy of the $J=1$ level, when 55\% of the molecules are found in it.
}
\label{fig:coanalysis}
\end{center}
\end{figure*}

In the left hand panel in Fig.~\ref{fig:coanalysis} we show the $\rm ^{12}CO / ^{13}CO$ brightness temperature ratio, $R_{12/13}$, as a function of the isotope ratio, $X_{12/13}$, and the optical depth, $\tau_{12}$, obtained directly from equation~\ref{eqn:opticaldepth}.  Essentially the brightness temperature ratio is constant, and equal to the isotope ratio, when the lines are optically thin.  It then rapidly drops (for a constant isotope ratio) as $\tau_{12}$ becomes optically thick, to reach the limit of $R_{12/13} = 1$ when both lines are optically thick.

The brightness temperature $T_A$ is given by
\begin{equation}
T_A = \frac{T_A^*}{\eta} = f [J(T_{ex}) - J(T_{\rm CMB})] (1 - e^{-\tau_{12}})
\label{eqn:temp}
\end{equation}
for a gas excitation temperature $T_{ex}$, where $\eta$ is the telescope efficiency, $f$ is the beam filling factor, $T_{\rm CMB}$ is the temperature of the 2.726\,K cosmic microwave background and $J(T) = T_1/[e^{T_1/T}-1]$ (with $T_1 = h \nu / k$ = 5.5\,K, where $\nu$ is the line frequency and $h$ and $k$ the well-known physical constants).  We show in the middle panel of Fig.~\ref{fig:coanalysis} the brightness temperature, $T_A$, as a function of the excitation temperature and the optical depth (and assuming a beam filling factor $f$ of unity).  When the emission is optically thin then $T_A$ is generally very much less than the excitation temperature, but for optically thick emission, when $T_{ex} \gtrsim 10$\,K, then we obtain $T_A \sim T_{ex}$; i.e.\ the brightness temperature yields the gas temperature directly.

Finally in the right hand panel of Fig.~\ref{fig:coanalysis} we show the fraction of the CO molecules that are found in the $J=1$ level, as a function of the excitation temperature, $T_{ex}$.  This is given by the Boltzmann equation
\begin{equation}
\frac{N_1}{N_{\rm CO}} = \frac{g_1}{Q(T_{ex})} e^{-T_1/T_{ex}}
\label{eqn:fraccolumn}
\end{equation}
where $N_1$ and $N_{\rm CO}$ are the column densities in the $J = 1$ level and all levels, respectively, $g_1 = 3$ is the level degeneracy, $T_1 = 5.5$\,K the energy above ground, and $Q(T) = 2 T / T_1$ is the partition function.  The fractional column density in the $J=1$ level peaks at 55\% when the excitation temperature equals the energy level, but is generally in the 40--20\% region for the $T = 10 - 30$\,K temperatures typical of most molecular gas. 

To calculate $N_1$ itself for each velocity bin, from standard 
molecular radiative transfer theory \citep[e.g.,][]{1999ApJ...517..209G} it can be shown that 
\begin{equation}
N_{1} = T_{A} \delta V \frac{8 \pi k \nu^{2}}{A h c^{3}} \frac{\tau_{12}}{1 - e^{-\tau_{12}}}
\label{eqn:n1}
\end{equation}
\noindent
where $c$ is the speed of light, $\nu$ the frequency of the transition, $A$ its radiative decay rate and $\delta V$ the channel velocity spacing.  The total column density is then obtained from summing over all the velocity channels.

Observationally, the measured integrated CO line flux is often converted into a total H$_2$ column density using the $\rm X_{CO}$ factor.  This is an empirically determined average conversion value between line flux and number of molecules per unit area in the telescope beam.  For instance, \citet{2001ApJ...547..792D}, from their Galactic plane survey, estimate that $\rm X_{CO}$ ranges from $\rm 1.8 \times 10^{20}$ for $|b| > 5^{\circ}$ to $\rm 2.7 \times 10^{20}$ for $|b| < 1^{\circ}$, in units of $\rm cm^{-2} \, K^{-1} \, km^{-1} \,s$.  Using equations \ref{eqn:fraccolumn} and \ref{eqn:n1} the $\rm X_{CO}$ factor can also readily be calculated given the [CO/H$_2$] abundance ratio, the CO optical depth and the excitation temperature, and substituting the relevant molecular parameters for CO\@.  This yields, when $T_{ex} = 10$\,K,

\begin{equation}
\rm
X_{CO} \sim 2.5 \times 10^{20} \left(\frac{3 \times 10^{-5}}{[CO/H_2]}\right) \left(\frac{\tau_{12}}{10}\right) cm^{-2} K^{-1} km^{-1} s.
\label{eqn:xfactor}
\end{equation}
Here we have set $\rm \tau_{12} = 10$ as our results (see \S\ref{sec:results} \& \S\ref{sec:discussratio}) show that this is a typical value in G323.
For $T_{ex} = 20, 40$ or 80\,K, $\rm X_{CO}$ in eqn.~\ref{eqn:xfactor} should be multiplied by 1.5, 2.6 or 4.9 times, respectively, as determined from the right hand panel of Fig.~\ref{fig:coanalysis}.  

For completeness, we also note that the preceding analysis applies equally well for the $\rm ^{12}CO / C^{18}O$ and the $\rm ^{13}CO / C^{18}O$ line ratios, as well as for the $\rm ^{13}CO$ and $\rm C^{18}O$ brightness temperatures, with corresponding use of the appropriate isotope ratios and optical depths in the relevant equations.

\subsection{Galactic Rotation and Source Distance}
\label{sec:galrot}
The $V_{LSR}$ radial velocities measured for any emission features seen provide estimates for their distances, $D$, from the Sun on the assumption that the Galactic rotation curve is known.  In this section we present calculations that allow these distances to be determined for the G323 field surveyed (and, by extension, for the full CO survey being undertaken).  A source radial velocity, as seen from our local standard of rest frame, is given by
\begin{equation}
V_{LSR} = V(R) \cos(\alpha) - V_0 \sin(l)
\end{equation}
where $V(R)$ is the source orbital speed about the centre of the Galaxy at a galactocentric radius $R$, with $V_0$ the orbital speed for the Sun. $l$ is the galactic longitude (taken to be 323.5$^{\circ}$ for the G323 field) and $\alpha$ is the angle measured, as seen from the Galactic centre, between the source and the tangential position along our sightline for the longitude $l$.  For the $4^{th}$ quadrant such velocities are negative within the Solar circle and positive beyond it. The orbital speed is obtained from the fit to HI data in the 4$^{th}$ quadrant derived by \citet{2007ApJ...671..427M}\footnote{i.e.\ $V(R) = V_0 [b_1 (R/R_0) + b_2]$ with $V_0 = 220$\,km\,s$^{-1}$, $R_0 = 8.5$\,kpc (the Sun's distance from the Galactic centre) and $b_1, b_2 = 0.186, 0.887$, respectively.  Note also that this yields an orbital velocity for the Sun of 236 km/s.}.  Beyond the Solar circle we use the \citet{1993A&A...275...67B} rotation curve, with the Sun's orbital velocity scaled to match that of the McClure-Griffiths and Dickey curve to avoid discontinuities at $R = R_0$.  For G323, however, we have only identified one clear positive velocity feature in the data set.  

\begin{figure*}[h]
\begin{center}
\includegraphics[scale=0.8, angle=0]{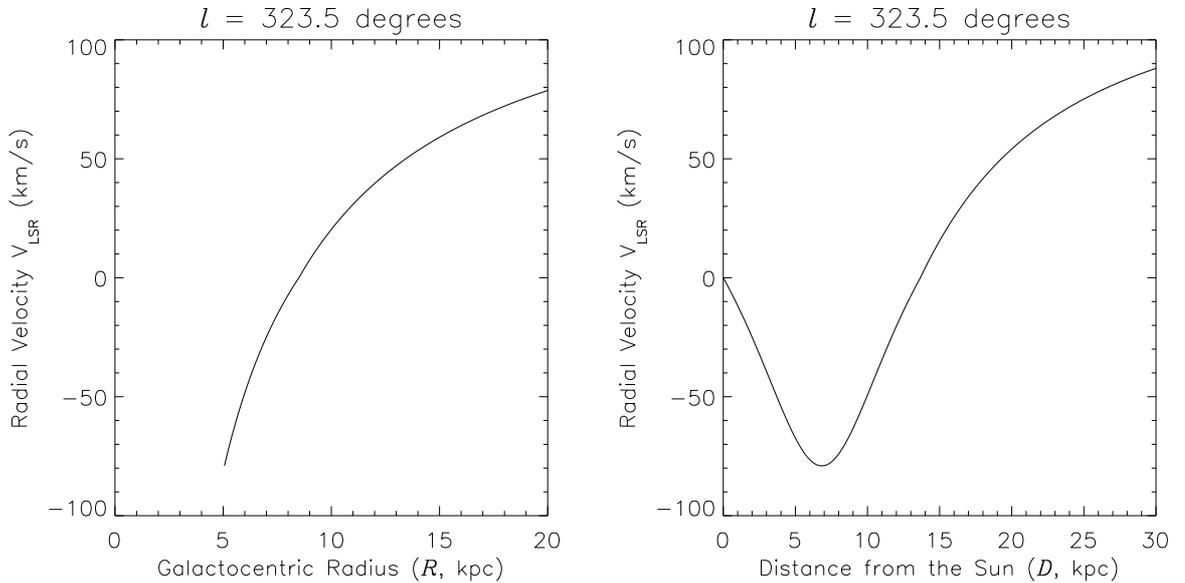}
\caption{Radial velocity -- distance relationships calculated for $l=323.5^{\circ}$ using the \citet{2007ApJ...671..427M} rotation curve for the inner Galaxy (i.e.\ negative velocities, with $R < R_{\odot}$) in the 4th quadrant, with the \citet{1993A&A...275...67B} curve for the outer Galaxy (i.e.\ positive velocities, and scaled to give the same orbital velocity at $R = R_{\odot}$).  To left the galactocentric radius in kpc is plotted against radial velocity, $V_{LSR}$ in km/s.  To right distance from the Sun, $D$ in kpc is plotted against $V_{LSR}$.  Near-distance solutions assume  $D < D_{tangent} = 6.8$\,kpc.  Far-distance solutions are for $D_{tangent} < D < 2\,D_{tangent}$.}
\label{fig:galrotation}
\end{center}
\end{figure*}

In Fig.~\ref{fig:galrotation} we show the derived relations between radial velocity and galactocentric radius ($R$, left) and distance from the Sun ($D$, right) for $l = 323.5^{\circ}$.  The latter, of course, displays the near-far ambiguity (i.e.\ the two Sun-distance solutions) for negative velocities between 0\,km/s and the tangent velocity along the sight line (of $-79$\,km/s). More negative velocities are `forbidden' in the sense that they are not possible under the assumption that the rotation curve holds.  This occurs at a tangential distance of $D = 6.8$\,kpc from the Sun (which is also $R = 5.1$\,kpc from the centre of the Galaxy).  For our analysis we thus assign such velocities to the tangential distance. Any positive velocity emission, on the other hand, would correspond to source distances greater than twice this; i.e.\ $D > 13.6$\,kpc.

\section{Results}
\label{sec:results}

In this section we present a selection of spectra and images that demonstrate some of the principal characteristics of the data set that the survey is yielding.  We calculate optical depths and column densities, and determine distances and masses for the emitting material.  We also examine how the $\rm ^{12}CO / ^{13}CO$ ratio and the optical depth vary along the sight line.

\subsection{Averaged Spectrum over the \\ G323 Survey Region}
\label{sec:average}
\begin{figure*}[h]
\begin{center}
\includegraphics[scale=0.9, angle=0]{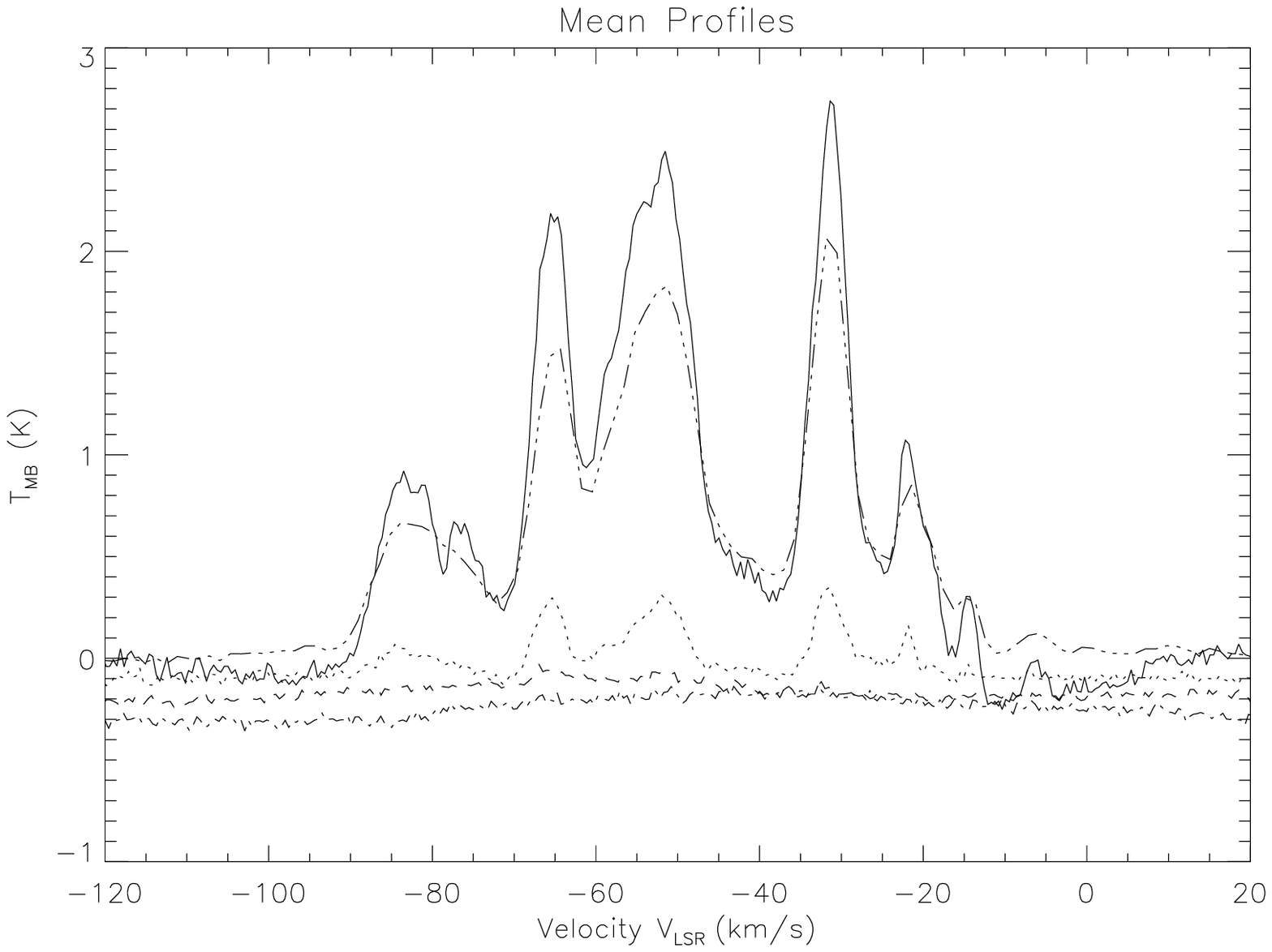}
\caption{The CO line profiles, averaged over the full G323 $1^{\circ}$ survey field, in units of $T_{MB}$ (K) (i.e.\ divided by the efficiency, $\eta_{XB} = 0.55$).  A binning of binning of 5 pixels (0.5 km/s) is used.   Shown from top to bottom (and each offset by $-0.1$\,K for clarity) are the $\rm ^{12}CO, ^{13}CO, C^{18}O \, \& \, C^{17}O$ lines.  For comparison the equivalent spectrum from the \citet{2001ApJ...547..792D} $^{12}$CO data cube is overlaid as a dot-dashed line.}
\label{fig:meanprofiles}
\end{center}
\end{figure*}

An averaged spectrum for the four CO lines measured, from the entire $1^{\circ}$ region surveyed in G323, is shown in Fig.~\ref{fig:meanprofiles}\footnote{Note also that these profiles have also been re-baselined.  The continuum level subtracted, now averaged over the entire $1^{\circ}$ survey field, provides an additional check on the accuracy of the baselining of the cube, which was carried out on a pixel-by-pixel basis.  It amounts to $\sim 10$\,mK.}.  This shows emission in the $\rm ^{12}CO$ line extending from $V_{LSR} \sim -5$ to $-90$\,km/s.  Several prominent spectral features are evident, with typical averaged peak brightness temperatures of $T_A^* \sim 1$\,K and widths of 5--10\,km/s (FWHM).  The $\rm ^{13}CO$ spectrum shows similar features, albeit with $10-20\%$ the intensity.  No features are evident in the averaged spectrum of either $\rm C^{18}O$ or $\rm C^{17}O$; however we do show below spectra showing the detection of the $\rm C^{18}O$ line at selected spatial locations.  The most blue-shifted feature, from $-80$ to $-90$\,km/s, in fact extends $\sim 10$\,km/s beyond the `forbidden' velocity limit of the adopted galactic rotation curve, though this is within typical values expected for non-circular orbital deviations.

\begin{figure*}[h]
\begin{center}
\includegraphics[scale=0.8, angle=0]{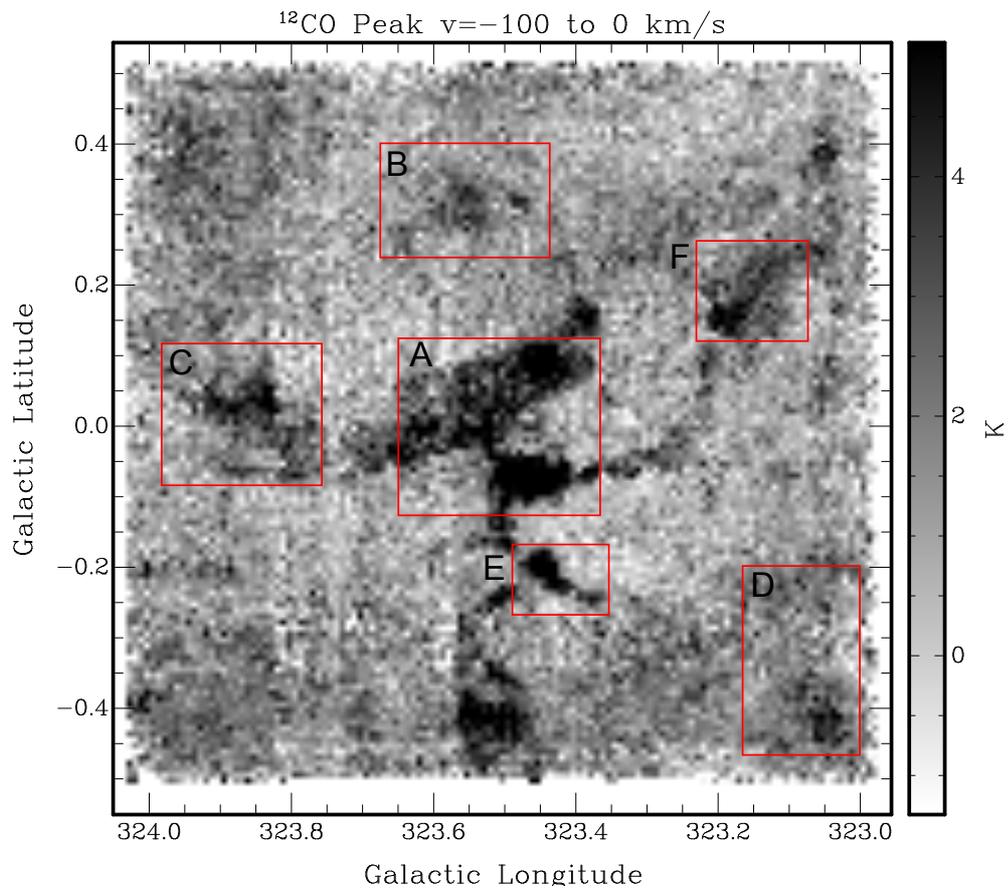}
\caption{Locations of the apertures defined in Table~\ref{tab:apertures}, overlaid on the $^{12}$CO peak temperature image ($T_A^*$ in K).  These need to be divided by the efficiency, $\eta = 0.55$, to yield main beam temperatures, $T_{MB}$.}
\label{fig:apertures}
\end{center}
\end{figure*}

\begin{figure*}[h]
\hspace*{0cm}
\includegraphics[scale=0.42, angle=0]{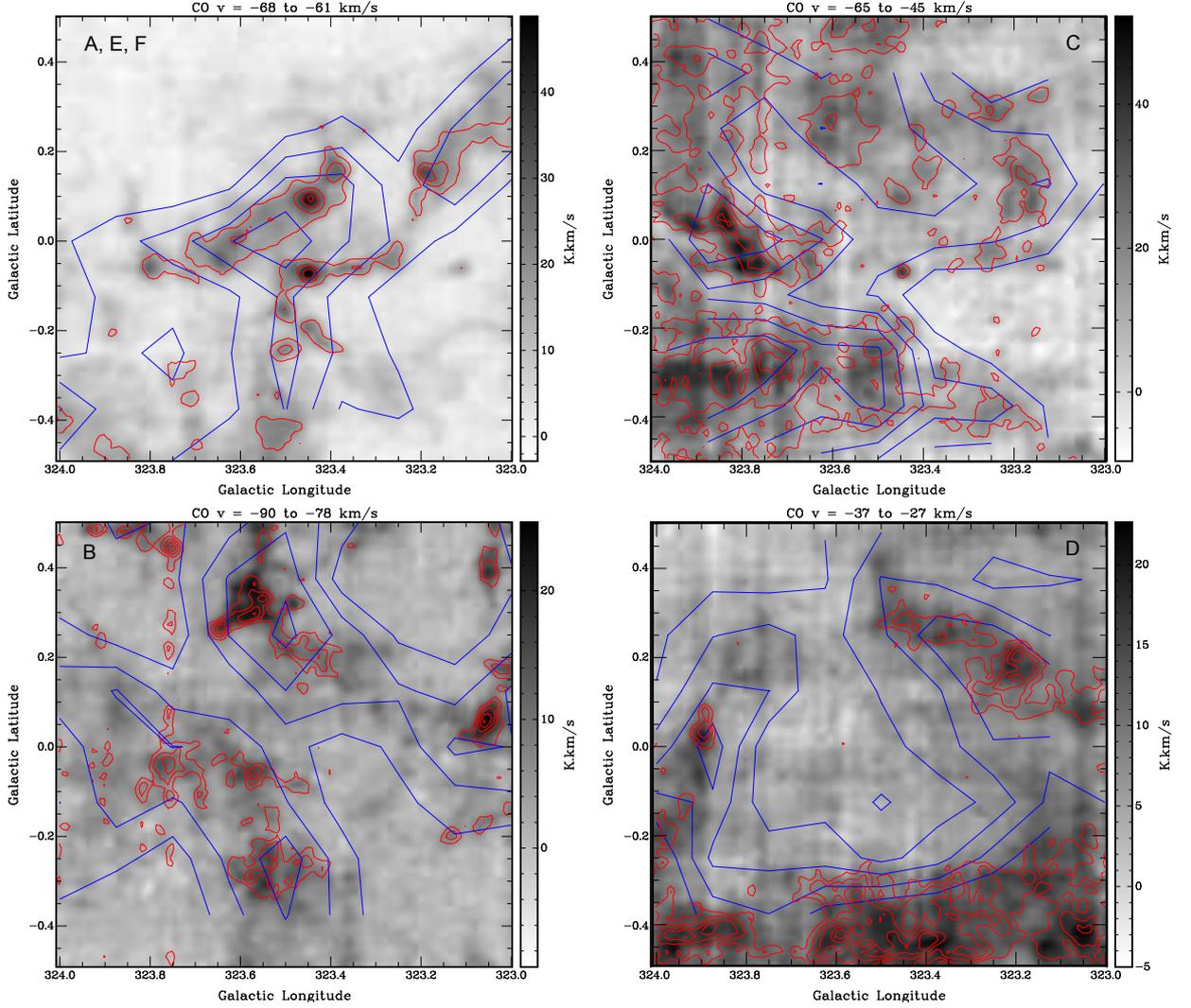}
\caption{Integrated flux images (in K km/s for $T_A^*$, as indicated by the scale bars) for the $\rm ^{12}CO$ line for the four different velocity ranges for the apertures listed in Table~\ref{tab:apertures}, overlaid with red contours of $\rm ^{13}CO$ line flux.  In blue are the corresponding contours obtained from the \citet{2001ApJ...547..792D} $^{12}$CO data cube.  From top-left, going clockwise, these velocity ranges are: $-68$ to $-61$, $-65$ to $-45$, $-37$ to $-27$ and $-90$ to $-78$\,km/s.   The letters (A -- F) relate to the relevant apertures.  Images have been smoothed with a $1'$ FWHM Gaussian beam.  $\rm ^{13}CO$ contour levels in the top two images are at 3, 6, 9 and 12\,K\,km/s, and in the bottom two at 2, 3, 4, 5 and 6\,K\,km/s. Note that some artefacts arising from the scanning directions used for OTF-mapping are apparent, as these are amplified when summing the data over many velocity channels.}
\label{fig:fluximages}
\end{figure*}

A peak intensity image for $\rm ^{12}CO$ line emission from the G323 region is shown in Fig.~\ref{fig:apertures}.  We note that for display purposes peak intensity images are generally superior than integrated intensity images over wide velocity ranges.  This is because residuals from imperfect baselining of rows or columns leave artefacts in integrated intensity images, even when not readily apparent in individual velocity channel images, as they contribute to every channel combined to form such an image.  While the magnitude of such residuals is not high, they are clearly apparent to the eye.  The analysis of the dataset (below), of course, uses the relevant integrated intensity images, as shown in Fig.~\ref{fig:fluximages}.

\subsection{Selected Apertures}
\label{sec:apertures}
Several apertures covering clearly identifiable features in the data cube were selected for further analysis.  These are also identified on the image in Fig.~\ref{fig:apertures}, as well as listed in Table~\ref{tab:apertures}.  Table~\ref{tab:apertureparams} then shows the adopted $V_{LSR}$ for these apertures (the mid-point of the velocity range), as well as the assumed distance and areal coverage on the sky, assuming the near-distance solution for the galactic rotation curve.  Further, an intrinsic $\rm [^{12}C / ^{13}C]$ isotope ratio, $X_{12/13}$, can be inferred given the derived galactocentric radius and its variation with this distance (we have applied $X_{12/13} = 5.5 R + 24.2$, where $R$ is the galactocentric radius in kpc; \citet{1982A&A...109..344H})\footnote{We note that there is considerable uncertainty, as well as scatter, in the value of the $\rm [^{12}C/^{13}C]$ isotope ratio as a function of galactocentric distance.  However, the resulting variation in the derived optical depth for our data set is small, and so this has little effect on the column densities we determine.}.  The isotope ratios we thus determined are also listed in Table~\ref{tab:apertureparams}.

\begin{figure*}[h]
\begin{center}
\includegraphics[scale=1.0, angle=0]{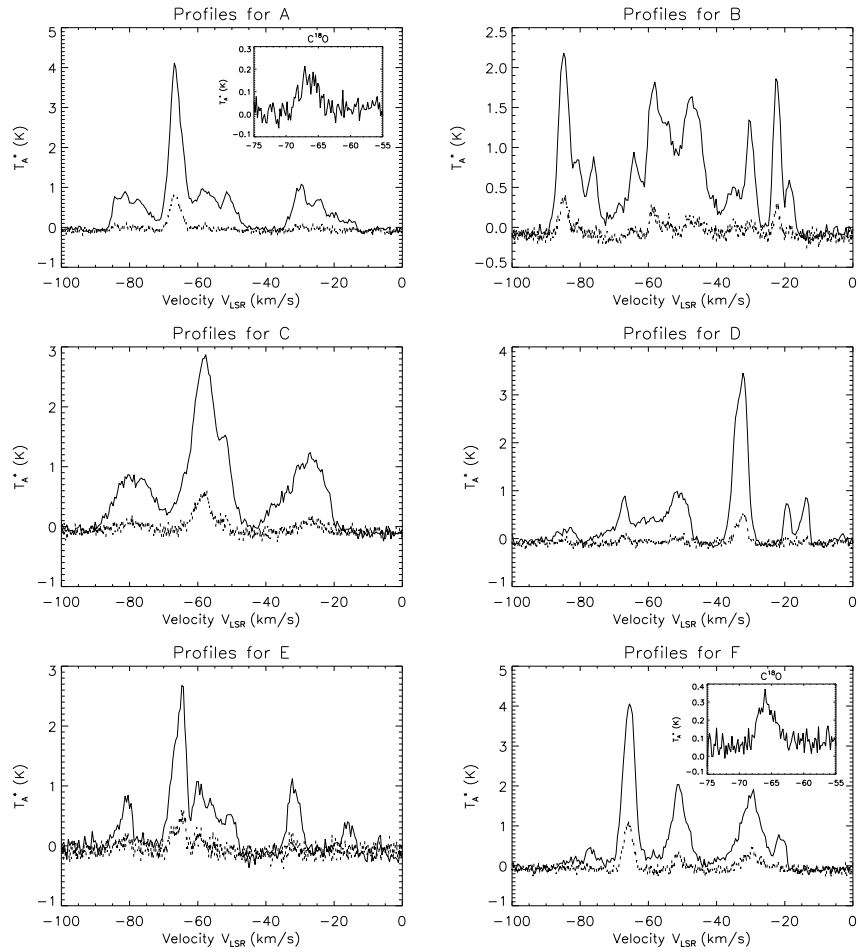}
\caption{The $\rm ^{12}CO$ (solid) and $\rm ^{13}CO$ (dashed) line profiles, averaged over each the 6 apertures specified in Table~\ref{tab:apertures}, in units of $T_A^*$ (K) (note that the velocity range shown is from $-100$ to 0\,km/s in each case and a binning of 5 channels (0.5 km/s) is used for the display).  For apertures A \& F, where the $\rm C^{18}O$ line is clearly detected, the inset also shows this line profile over the aperture's velocity range.  }
\label{fig:apertureprofiles}
\end{center}
\end{figure*}

Figure~\ref{fig:apertureprofiles} presents the $\rm ^{12}CO$ and $\rm ^{13}CO$ profiles for these apertures and in Fig.~\ref{fig:fluximages} their $\rm ^{12}CO$ line flux images, overlaid with $\rm ^{13}CO$ contours (and the Dame et al.\ $^{12}$CO contours -- see \S\ref{sec:comparison}).  Table~\ref{tab:linefluxes} lists the integrated fluxes for the 4 CO lines, for each of the apertures defined in Table~\ref{tab:apertures}, together with their errors.  These errors include both the statistical error, determined from the standard deviation of the data in continuum portion of each spectrum, and an estimate of the error in determining the level of continuum itself.  In general, the latter is the dominant source of error for the integrated line fluxes\footnote{For instance, over the 7\,km/s velocity width used for apertures B, E \& F, the statistical error on the line flux is approximately 15\% of the error we determine arising from the uncertainty of the continuum level.}.

Line ratios for $\rm ^{12}CO/^{13}CO$ in these apertures are then listed in Table~\ref{tab:lineratios}, together with the calculated optical depths, $\tau_{12}$, of the $\rm ^{12}CO$ line (the latter also assuming the isotope ratio listed in Table~\ref{tab:apertureparams}).  These quantities vary little between the apertures, ranging from $\sim 5 - 10$ in both cases.

\subsection{C$^{18}$O and Positive Velocities}
\label{sec:ununsual}
Only two clear, and one marginal, detection of the $\rm C^{18}O$ line are seen in the apertures (A, F \& C, respectively).  The clear detections are shown as insets in Fig.~\ref{fig:apertureprofiles}. The (optically) thin ratio $\rm ^{13}CO / C^{18}O$ is found to be approximately equal to 4 for these apertures.  This is somewhat smaller than the ratio determined in the G333 molecular cloud of $\sim 6$ by \citet{2008MNRAS.386.1069W}, or indeed the  abundance ratio of 7.4 adopted by the same authors; however our determination of the ratio is not sufficiently precise to test whether this amounts to a real difference.

We note that only one clear (but weak) detection is seen at positive velocities in the data set.  It is centred at $l=323.565^{\circ}, b=0.250^{\circ}, V_{LSR}=8.2$\,km s$^{-1}$.  This places the feature 14.3\,kpc away. The integrated line flux, averaged over a $\rm 0.05^{\circ} \times 0.06^{\circ} \times 4.5 \, km \, s^{-1}$ aperture, is $\rm 9 \pm 2 \, K \, km \, s^{-1}$.

\subsection{Pencil Beams}
\label{sec:pencil}
\begin{figure*}[h]
\begin{center}
\includegraphics[scale=1.0, angle=0]{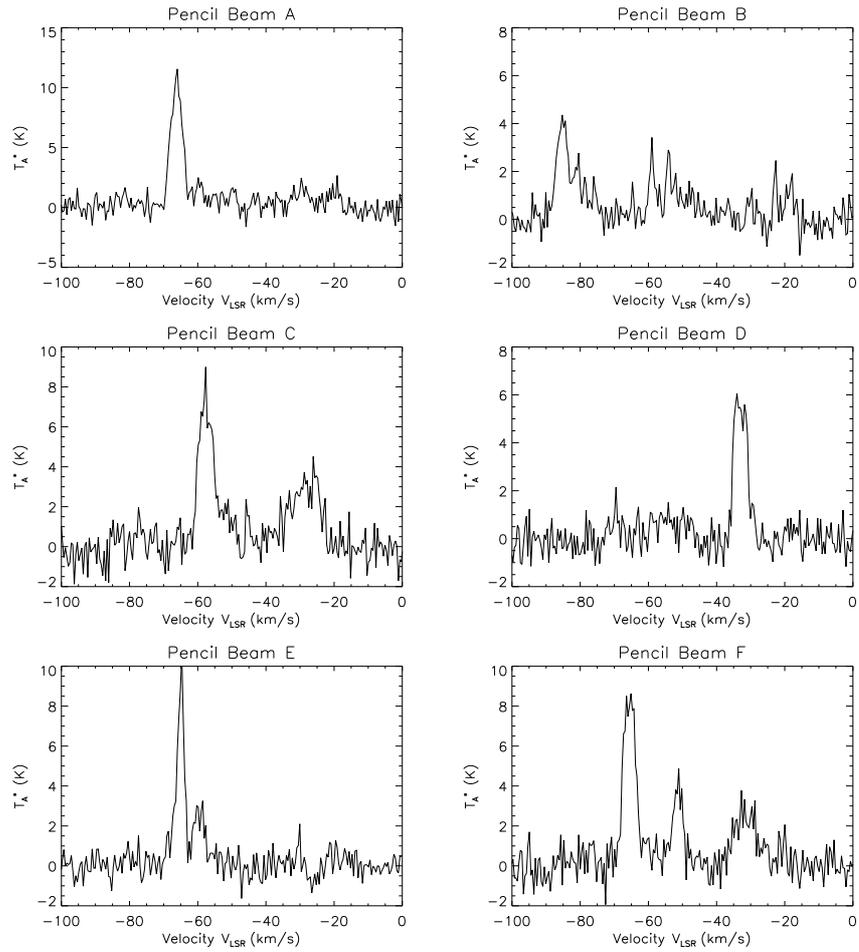}
\caption{The $\rm ^{12}CO$ line profiles at the peak pixel position within each of the 6 apertures specified in Table~\ref{tab:apertures}, in units of $T_A^*$ (K).  A binning of 5 channels (0.5 km/s) is used for the display.}
\label{fig:pencilbeams}
\end{center}
\end{figure*}

Table \ref{tab:linefluxes} also tabulates the peak $^{12}$CO line channel brightness measured within each aperture (and the corresponding $^{13}$CO brightness for this pixel), with the spectra shown in Fig.~\ref{fig:pencilbeams}.  The highest brightness temperature in the G323 field is $\sim 13$\,K, equivalent to $T_{MB} \sim 30$\,K\@.\footnote{As per the analysis in \S\ref{sec:calcs} for middle panel of Fig.~\ref{fig:coanalysis} and applying $\eta_{MB} = 0.42$ as the main beam, rather than the extended beam, efficiency for the pencil beam.}  If the gas is filling the beam this would be equal to the gas temperature at the position where the line becomes optically thick along this sight line (and a lower limit if not). Comparison to Fig.~\ref{fig:apertureprofiles} also provides an indication as to how the CO line flux is distributed over the apertures.  The peak line brightness is typically 3--4 times larger for these $30''$ pencil beams than when averaged over the aperture.  The $^{12}$CO/$^{13}$CO line ratio is, however, 2--3 times larger for the integrated fluxes than for the peak pixel (where it is $\sim 2 - 3$).  This indicates that the optical depth of $^{12}$CO emission at the emission peaks are 2--3 times larger than the mean value for the apertures.  

\subsection{Column Densities and Masses}
\label{sec:column}
Finally, column densities and molecular masses are presented in Table~\ref{tab:columndensity}.  We present two estimates here for the column density.  The first estimate makes use of the empirical $\rm X_{CO}$ X--factor derived by \citet{2001ApJ...547..792D} (see \S\ref{sec:calcs}).  The second applies a radiative transfer calculation to the level population distribution (i.e.\ using equation \ref{eqn:xfactor}, derived from eqns.\ \ref{eqn:fraccolumn} and \ref{eqn:n1}), assuming $T_{ex} = 10$\,K and a [CO/H$_2$] abundance of $3 \times 10^{-5,}$\footnote{These later values can be scaled for different abundances and temperatures, as discussed in \S\ref{sec:calcs}.}.  The two different estimates are comparable for this choice of parameter values, as is evident from eqn.~\ref{eqn:xfactor}.  Typical column densities for $N_{H_2}$ of $\rm  6-8 \times 10^{21} \, cm^{-2}$ for the apertures A--F are obtained.  This is consistent with the average column density determined for GMCs (e.g.\ of $N_{H_2} =  8 \times 10^{21}$\,cm$^{-2}$; \citet{1987ApJ...319..730S}).  For the complete survey field, integrating across the full $\rm 100 \, km\,s^{-1}$ velocity range of the emission, an $N_{H_2}$ of $\rm 2 \times 10^{22} \, cm^{-2}$ is obtained, equivalent to an optical depth $A_v \sim 20$ magnitudes.  Similar column densities are also obtained for the complete sight-lines along each of the apertures A--F\@. They are equivalent to an average number density $n_{H2} \sim 1$\,cm$^{-3}$ over $\sim 7$\,kpc distance to the tangent point along the sight-line.

Furthermore, for the pencil beams (Fig.~\ref{fig:pencilbeams}) the greater brightness and optical depth at the peak pixels implies that the corresponding column densities  are an order of magnitude higher than the average column density over the apertures.  Peak column densities along these sight lines thus reach $N_{H_2} \sim 10^{23}$ $\rm cm^{-2}$.

Masses can also be derived for each feature, given a distance to them (and their spatial and velocity extents).  We assume the near-distance solution for the line $V_{LSR}$ and so obtain the masses listed in Table~\ref{tab:columndensity} when applying the X--factor calculation.  Masses from the radiative transfer calculation of the column density can readily be obtained by appropriate scaling of the values listed in Table~\ref{tab:columndensity} by the ratios of the column densities determined  (and also for different assumed temperatures by using the right hand plot in Fig.~\ref{fig:coanalysis}).  

The masses for these selected features are found to typically be of order $10^{4} \, M_{\odot}$.  This amounts to a few percent of the total molecular mass within the survey region (see below).  Indeed, summing over over the masses of the 20 brightest features readily identifiable in the data cube yields $\sim 20\%$ of the total mass in the 1 square degree aperture.  This indicates that most of the molecular mass in region is well distributed between many emitting clumps, and is not dominated by just the few brightest features.

\subsection{Position -- Velocity Plots}
\label{sec:pvplots}
\begin{figure*}[h]
\includegraphics[scale=0.4, angle=-90]{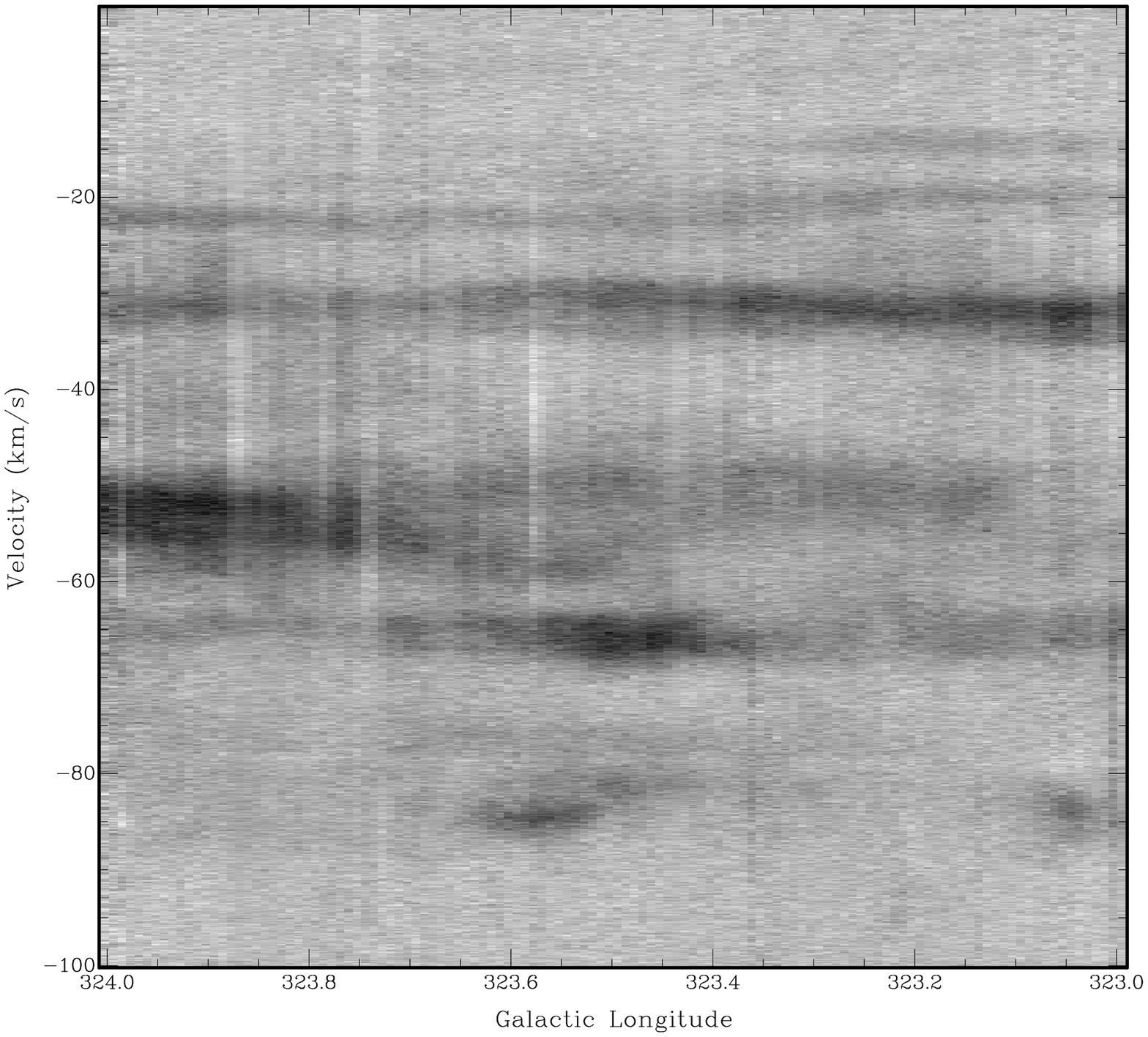}
\includegraphics[scale=0.4, angle=-90]{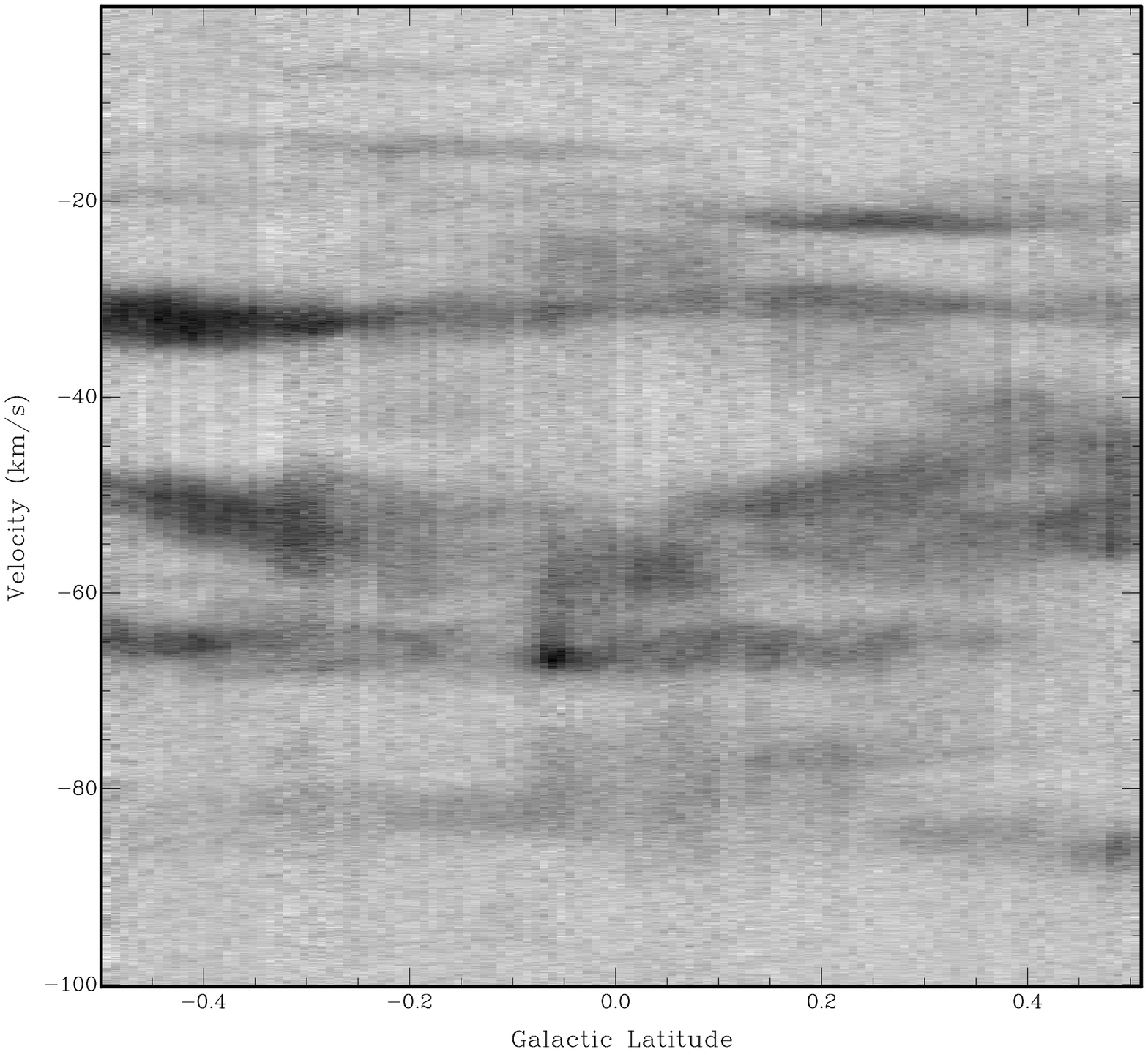}
\caption{Position--velocity images from the G323 data cube for the $\rm ^{12}CO$ line.  To left, the $V_{LSR}$ radial velocity in km/s is plotted on the $y$-axis against Galactic longitude, $l$ on the $x$-axis.  To right, it is plotted against Galactic latitude, $b$.  In each case the data from the other direction ($b, l$, respectively) has been averaged over the entire degree covered by the data cube.  Note that some residuals from poor data are evident in the striping structure seen in the velocity direction.  Several bands are seen running across the images at roughly constant velocities.  These may be related to spiral arms crossed along the sight line through the Galaxy, as discussed in \S\ref{sec:pvplots}. }
\label{fig:pvimages}
\end{figure*}

The final figure we present in this section shows two position-velocity slices (`PV--images') in Fig.~\ref{fig:pvimages}.  These plot the $^{12}$CO emission brightness as a function of radial velocity against either Galactic longitude (left) or latitude (right) over the $1^{\circ}$ G323 region surveyed, with the data averaged over the other spatial direction (i.e.\ $b$ and $l$, respectively).  Clearly evident are several bands running across the PV--images at roughly constant velocities.  These may be related to molecular clouds associated with spiral arms traversed along the $l=323^{\circ}$ sight line through the Galaxy.  While the identification with specific spiral arms in, e.g.\ the Galactic model presented by \citet{2008AJ....135.1301V} (whose nomenclature for the arms we adopt below) may be ambiguous with just the $1^{\circ}$ of longitude coverage presented here, comparison of our data with the lower resolution data of \citet{2001ApJ...547..792D} allows most of the features to be identified with specific spiral arms. 

The dominant feature is the passage through the near-distance portion of the Scutum--Crux arm 3--4\,kpc away, producing the emission seen in the $-50$ to $-65$ km/s range.  The most blue-shifted emission, around $-80$ to $-90$\,km/s, can be associated with the tangent to the Norma--Cygnus arm, around 7\,kpc distant (we also note that this actually exceeds by $\sim 10$\,km/s the forbidden velocity limit given by the adopted Galactic rotation curve, as discussed in \S\ref{sec:galrot}). At low radial velocities, around $-20$\,km/s, emission from the near-distance to the Sagittarius--Carina arm is seen, $\sim 2$\,kpc away.  It is possible that the strong emission band at $-35$\,km/s also arises from this arm, but it may also come from the Scutum--Crux arm; the emission velocity is mid-way between that expected for either of these arms.  A weak band of emission is also seen at $-8$\,km/s.  This is most likely associated with the far-distance portion of the Scutum-Crux arm ($\sim 13$\,kpc away), rather than the much nearer Sagittarius-Carina arm.  

\subsection{Comparison with the \\ Dame et al. Survey}
\label{sec:comparison}
In this section we briefly compare the results of our survey with the $^{12}$CO data cubes available from the lower resolution survey conducted by \citet{2001ApJ...547..792D} (as shown in Fig.~\ref{fig:survey}).  In Fig.~\ref{fig:meanprofiles} their line profile, averaged over the entire $1^{\circ}$ survey field, is overlaid on the corresponding profile from our data cube.  The structure of the profiles match well, albeit at their lower spectral resolution ($\rm 1.3\, km\,s^{-1}$).  The intensity of the Mopra  line peaks, however, are typically $\sim 30 - 50\%$ higher than the corresponding peaks in Dame et al.\ data, although the minimums between the peaks are roughly similar in both.

Line morphologies are compared in Fig.~\ref{fig:fluximages}, where the Dame et al.\ integrated flux maps are overlaid on the corresponding images from our data cube for the relevant velocity ranges for the apertures.  While the difference in angular resolution is obvious ($9'$ vs.\ $0.5'$) the morphology of the emission regions are clearly similar.  The Mopra fluxes are, however, again higher than those determined from Dame et al., by between $20 - 100$\% (the Dame et al. fluxes are listed in the last column of Table~\ref{tab:linefluxes}).   However these apertures correspond to only a few $9'$ pixels in the Dame et al.\ data, so  our apertures cannot be well-matched on the sky, with typically a larger effective area on the sky included in the calculation of fluxes for the Dame et al.\ data set.  The comparison for the flux from the entire 1$^{\circ}$ survey region, where this issue is minimal, is closer, with the Mopra value being $\sim 30\%$ higher.  On the other hand, the error on the continuum determination over the 100\,km/s wide profiles for this latter case is also larger (for instance, a 50\,mK offset corresponds to an uncertainty in the line flux of 5\,K\,km/s); if we do not fit for a continuum level for the $1^{\circ}$ aperture then the Mopra line flux is only 10\% higher.

This indicates that, while the spatial and spectral structures in the data cubes between our two surveys are clearly consistent with each other, that there is a difference in their absolute calibration.  A possible cause is an under-estimation of the beam efficiency of the Mopra data set (and/or an over-estimation of that of Dame et al.), as all intensities are scaled by these assumed values (0.55 and 0.82, respectively).  If our flux calibration is too high then this would imply smaller column densities (and masses) than those we list in the Tables by the relevant calibration factor fractional error.  We conclude that the flux scale between these two data sets may differ by up to $\sim 30\%$.
As discussed in \S\ref{sec:observations}, our  monitoring of the source M17 SW suggests, however, that the absolute flux scale for Mopra is similar to that of the SEST telescope at 115\,GHz.

\section{Discussion}
\label{sec:discussion}
\subsection{Line Ratio and Optical Depth Variations}
\label{sec:discussratio}
The data set may also be used to examine the variation of the $\rm ^{12}CO/^{13}CO$ line ratio and the $\rm ^{12}CO$ optical depth along the sight line.  We do this for an averaged aperture covering the entire $1^{\circ}$ region surveyed in G323\@.  We also make the assumption that the $V_{LSR}$ velocities can each be identified with features at the near distance rather than the far distance.  This is clearly a simplification, but the discussion in \S\ref{sec:pvplots}, based upon identification of the spectral features with a model for the Galaxy, suggests that it is a good approximation for this region of the Galaxy.  We also examine this assumption further in \S\ref{sec:hiabsorption} based on a comparison with the HI data from the SGPS \citep{2005ApJS..158..178M}.  A more comprehensive study over several degrees of longitude may be able to discern which features are associated with near- and far-distances based on comparison with the known positions of spiral arms (e.g.\ as identified by \citet{2008AJ....135.1301V}).  However, the error we make through this assumption is relatively small since the CO emission does not fill the $1^{\circ}$ aperture.  Hence the strength of any emission arising from the far-distance is reduced by the square of the distance ratio of far-to-near, in comparison to corresponding emission arising from the near-distance.  For features with $V_{LSR}$ near rest this is indeed small; e.g.\ for $V_{LSR} = -20$\,km/s the far-distance is 12\,kpc compared to 2\,kpc for the near-distance, a flux reduction of a factor of 36.  For $V_{LSR} = -50$\,km/s, where the bulk of the emission arises, the far-to-near distance ratio of $\sim 3$ corresponds to a reduction in flux of a factor of $\sim 9$.  For the emission nearest the tangent point our assumption is weakest; e.g.\ $-75$\,km/s yields a far-to-near ratio of $\sim 1.5$ and so a flux reduction of just $\sim 2$ times.  However, in this case this distance error is then relatively small.  Furthermore, as was discussed in \S\ref{sec:pvplots}, only the weak feature seen at $\sim -8$\,km/s is actually likely to be associated with a far-distance solution.  If so, then the features located at $\sim 1$\,kpc in the analysis below would actually be $\sim 13$\,kpc away for the adopted rotation curve.

\begin{figure*}[h]
\begin{center}
\includegraphics[scale=0.8, angle=0]{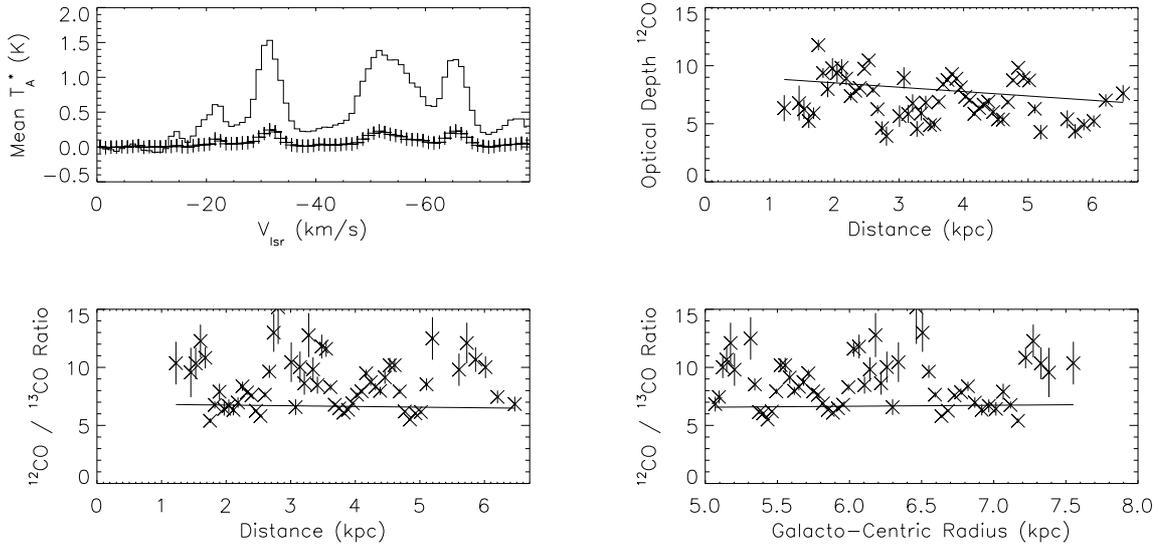}
\caption{Top-Left: mean brightness temperature ($T_A^*$ in K) of the $^{12}$CO line over the entire $1^{\circ}$ aperture, plotted as a function of radial velocity, $V_{LSR}$ between 0 and the tangent velocity ($-79$\,km/s), and binned into 1\,km/s intervals.  The + signs show the corresponding $^{13}$CO line brightness temperature.  Bottom-left: $R_{12/13} = \rm [^{12}CO / ^{13}CO]$ line ratio as a function of distance, $D$ in kpc, from the Sun assuming the $V_{LSR}$ velocities derive from the near-distance.  In each 1 km/s velocity bin the SNR for both lines must be $> 5$ for a ratio to be calculated; the corresponding $1 \sigma$ error bars are shown on the plot.  Bottom-right: as for bottom-left, except that galactocentric radius, $R_{Gal}$, is plotted against the $\rm ^{12}CO / ^{13}CO$ line ratio.  Note that there is no near-far ambiguity for this plot.  Top-right: optical-depth in the $\rm ^{12}CO$ line, as a function of distance from the Sun, also assuming a $\rm [^{12}C/^{13}C]$ isotope ratio given by $5.5\,R_{Gal} + 24.2$, where $R_{Gal}$ is in kpc \citep{1982A&A...109..344H}. In the latter three cases the straight lines are the best linear fits to the (error-weighted) data sets, and are given by (i) $R_{12/13} = (6.9\pm0.1)\,D -0.06\pm0.03$, (ii) $R_{12/13} = (6.1\pm1.4)\,R_{Gal} + 0.09\pm0.06$ and (iii) $\tau_{12} = (9.3\pm0.2)\,D -0.38\pm0.04$, respectively.  }
\label{fig:radialvariations}
\end{center}
\end{figure*}

We have binned the data into 1\,km/s velocity bins for this analysis (i.e.\ approximately 10 resolution elements), to improve the signal to noise ratio (SNR).  This yields a distance step along the sight line of $\sim 100$\,pc.  The corresponding variation of the $\rm ^{12}CO$ and $\rm ^{13}CO$ fluxes with radial velocity is shown in the top-left of the 4 plots in Fig.~\ref{fig:radialvariations}.  This binned data is then used to yield the other three plots shown in this Figure.  Furthermore, the SNR in both these lines is required to exceed 5 to be included in this analysis. At bottom-left the $\rm ^{12}CO / ^{13}CO$ line ratio is shown as a function of distance from the Sun, and at bottom-right as a function of galactocentric radius.  The distribution is flat, though clearly there is also significant scatter.  The lines overplotted in Fig.~\ref{fig:radialvariations} show the best linear fits; the mean ratio is found to be $\sim 7$.  No significant trend with distance from the Galactic centre is discernible, as for instance might have been expected for a gradient in the $\rm ^{12}C/^{13}C$ isotope abundance ratio.  However, the range in galactocentric radius sampled is also relatively small; from 5.5 to 7\,kpc.  Data from the full survey, which will cover a much wider range of galactocentric distances, will be better able to probe for any such variation.  Finally, the top-right panel of Fig.~\ref{fig:radialvariations} shows the derived optical depth, $\tau_{12}$, as a function of distance from the Sun.  Again, no significant variation is found, with the mean value being $\tau_{12} \sim 9$, as indicated by the best linear fit line. This determination also assumes the relevant value for the $\rm ^{12}C/^{13}C$ isotope ratio for each galactocentric radius, as discussed earlier in \S\ref{sec:apertures}.

\subsection{Column Density and Mass Distribution}
\label{sec:massdistribution}

\begin{figure*}[h]
\begin{center}
\includegraphics[scale=0.8, angle=0]{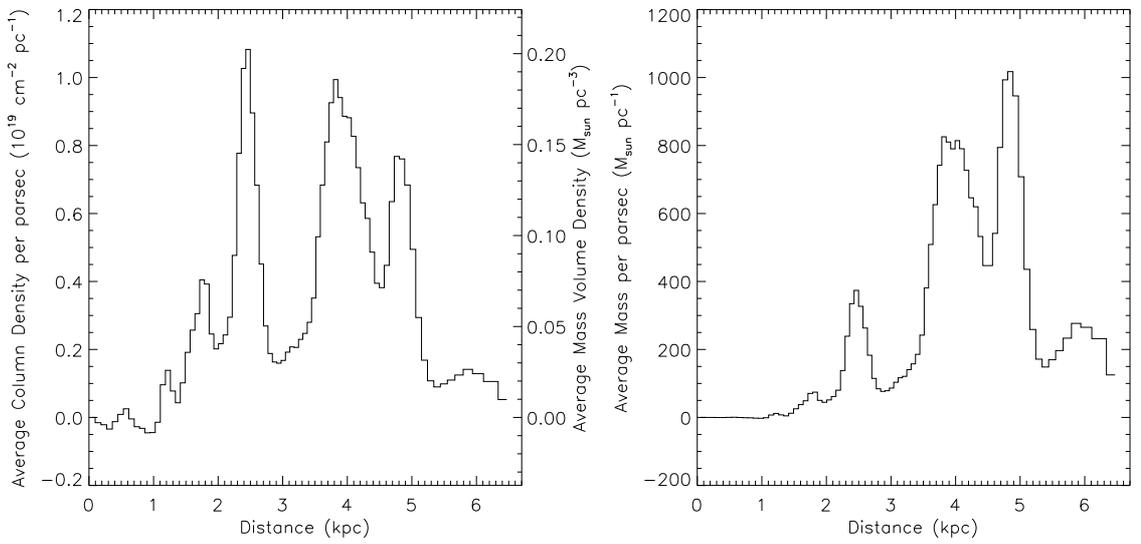}
\caption{Column density (left panel) and mass (right panel) per parsec along the sight line for the data set, integrated over the entire $1^{\circ}$ aperture of the G323 region surveyed.  Distances from the Sun, in kpc, assume the near-solution for the radial velocity and the data has been averaged over 1\,km/s bins.  In the left panel the column density per parsec, in units of $\rm 10^{19} \, cm^{-2} \, pc^{-1}$, is shown on the left-hand axis, and as average mass density in $\rm M_{\odot} \, pc^{-3}$ on the right-hand axis.  In the right panel the averaged mass per parsec ($\rm M_{\odot} \, pc^{-1}$) is shown; i.e.\ calculating the total mass per unit distance over the full 1 square degree area surveyed, for each distance from the Sun along the sight line. Larger distances, of course, encompass larger physical areas on the sky, so contributing to their, in general, greater masses.  The total integrated mass enclosed within the survey region is $\rm \sim 1.7 \times 10^6 M_{\odot}$.}
\label{fig:radialcut}
\end{center}
\end{figure*}

We now take the radial distribution of the line intensities shown in Fig.~\ref{fig:radialvariations} to examine the mass distribution of molecular gas with distance, under the same assumption of taking the near-distance solution for the radial velocities between 0 and $-79$\,km/s (the tangent velocity).  As described in \S\ref{sec:calcs}, the line fluxes in each 1\,km/s velocity bin yield column densities, and these are shown per unit distance along the sight line in the left hand panel of Fig.~\ref{fig:radialcut}.  The distribution essentially follows that of the line intensity, with the radial velocity now converted to a distance.  Half-a-dozen clear enhancements in column density are seen, typically separated by $\sim 1$\,kpc and each extending for $\sim 300$\,pc.  They correspond to crossings of the spiral arms. The column densities in these features are of order $\rm 10^{19} \, cm^{-2}$ per parsec, yielding total H$_2$ columns of several $\rm \times 10^{21} \, cm^{-2}$ per feature.  The same data can also be displayed as a mass density, calculating the mass per square parsec at each distance along the sight line.  This is shown as the right-hand axis of the same plot.  The same emission features yield average mass densities of $\rm \sim 0.1\,M_{\odot} \, pc^{-3}$ or $n_{H_2} \sim 1$\,cm$^{-3}$.

The total mass at each distance along the sight line can also be determined, given the area of the beam on the sky at each distance.  This then yields the mass per parsec along the sight line, as shown in the right hand panel of Fig.~\ref{fig:radialcut}.  This plot displays an increase in mass with distance, though it must be remembered that this is, in large part, caused by the increased physical size of the emitting region being probed by the survey at the further distances.  Nevertheless, it shows that mass linear densities of a few $\rm \times 10^2 \, M_{\odot}$ per parsec are found, equivalent to a total of a few $\rm \times 10^{5} \, M_{\odot}$ for the (four) principal molecular emitting regions crossed along the sightline.  This molecular gas is spread out over projected areas of $\sim 70 \times 70$\,pc for each of these emitting regions. In turn, this implies several tens of clumps contributing to the emission within each of these given the $\rm \sim 10^4  \, M_{\odot}$ typical mass of a clump (determined in \S\ref{sec:column}).  The total molecular mass enclosed within this $1^{\circ}$ aperture is found to be $\rm \sim 2 \times 10^6 M_{\odot}$ on integrating along the whole sightline through the Galaxy.  This is consistent with the total molecular gas for the Galaxy being some two orders of magnitude higher (i.e.\ several $\rm \times 10^8 M_{\odot}$) if we assume the $1^{\circ}$ G323 survey block is typical of the inner $\sim 100^{\circ}$ of the Galaxy, where the bulk of the molecular gas is found.  

If the feature at $-8$\,km/s  were actually at $\sim 13$\,kpc, as discussed above, rather than $\sim 1$\,kpc assumed, then its associated linear mass density would be two orders of magnitude greater.  However the column density, averaged over the (much larger) aperture size at that distance, is unchanged.  The total mass would be $\rm \sim 0.5 \times 10^6 \, M_{\odot}$ at this larger distance, so increasing the total mass budget by about one-quarter along the G323 sightline. 

\subsection{Distance Ambiguity and HI Self-absorption Features}
\label{sec:hiabsorption}
\begin{figure*}[h]
\begin{center}
\includegraphics[scale=0.9, angle=0]{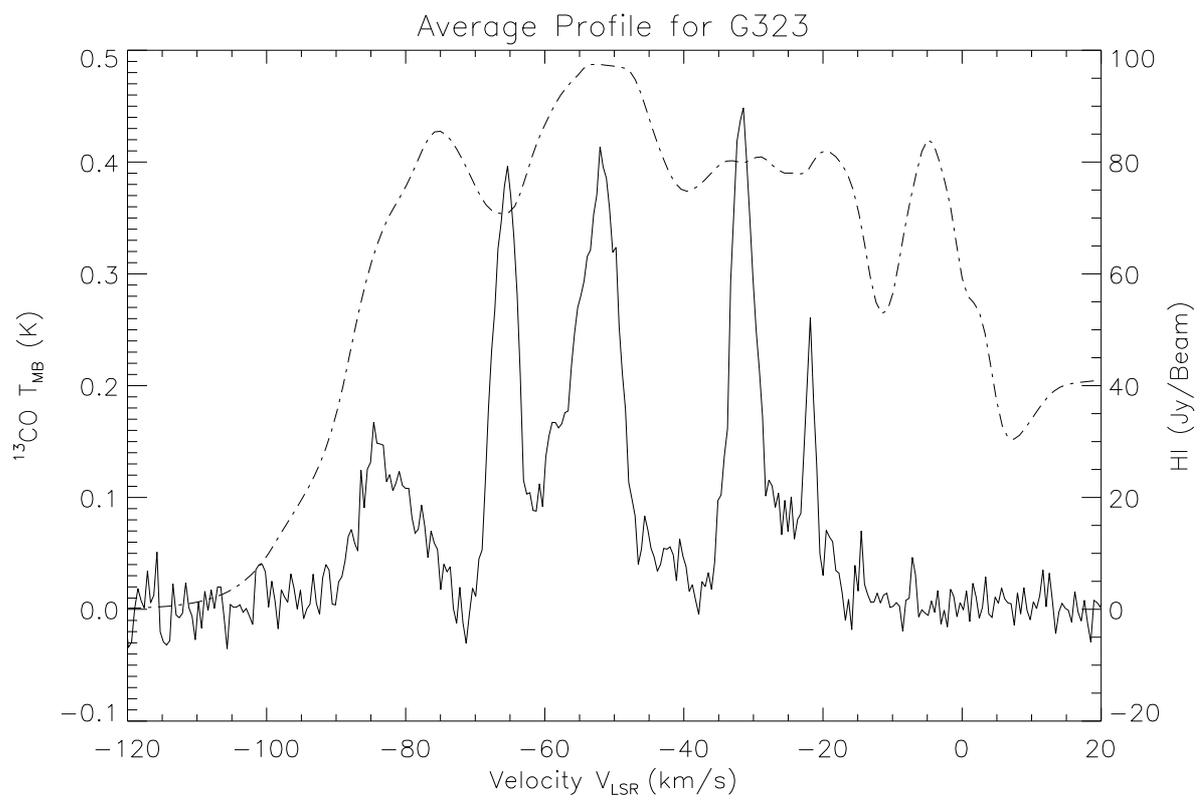}
\caption{Averaged profiles for the G323 $1^\circ$ survey region for the $^{13}$CO (this work) and 21\,cm HI \citep{2005ApJS..158..178M} lines.  The CO flux scale is on the left and the HI scale on the right axis. }
\label{fig:cohiprofile}
\end{center}
\end{figure*}

In this section we examine further our assumption that the majority of the CO spectral features measured in the G323 region arise from the near distance rather than the far distance solution of the Galactic rotation curve equation (as calculated in \S\ref{sec:galrot}).  As discussed by \citet{2002ApJ...566L..81J}, HI self-absorption associated with $^{13}$CO emission features may be used to indicate a near-distance solution.  This is based on the assumption that molecular clouds are surrounded by a cold atomic sheath that absorbs HI emission from warmer gas that is further away.  If this is seen then the cloud is more likely to be associated with the near-distance, rather than the far-distance, solution.  We make use of the publicly available Southern Galactic Plane Survey (SGPS) data set \citep{2005ApJS..158..178M} for this purpose, binning the Mopra $^{13}$CO data cube to the same voxel scale as the SGPS\footnote{Note that (optically thin) $^{13}$CO is used to minimise any possible comparison of HI with $^{12}$CO self-absorbed features.}.  As per the analysis of \citet{2002ApJ...566L..81J} we compared (i) the integrated intensity maps of these lines over the same velocity ranges, (ii) the integrated intensity profiles over the apertures in Table \ref{tab:apertures} and (iii) the scatter plots of HI {\it vs.}\ CO intensities per voxel.  However, for the sake of brevity, we only present here (see Fig.~\ref{fig:cohiprofile}) a plot of the integrated line profiles across the entire $1^{\circ}$ aperture.  This shows most of the relevant features we discuss below.  In all cases, aside from the weak $-8$\,km\,s$^{-1}$ feature, the balance of the evidence is consistent with a near-distance solution.

We now discuss each the features in turn.  For the apertures A, E \& F ($-65$\,km\,s$^{-1}$) in Table \ref{tab:apertures} there is a clear HI absorption dip at the same velocity of the CO feature in the integrated spectrum as well as in those for each aperture.  We also note that the scatter plot of the HI {\it vs.}\ CO intensities for each voxel in aperture A shows an anti-correlation between the lines, consistent with HI self-absorption at the brightest CO positions.  Aperture B ($-84$\,km\,s$^{-1}$) does not show any HI absorption feature, however it is associated with the tangent-point velocity and so does not have a near-far ambiguity.  Aperture C ($-55$\,km\,s$^{-1}$) does not show any HI absorption dip in the line profile, however the HI emission map shows a clear `hole' at the location of the CO peak.  Aperture D ($-32$\,km\,s$^{-1}$) shows a weak dip in the integrated HI profile.   The feature at $-22$\,km\,s$^{-1}$, discussed in \S\ref{sec:pvplots} as likely arising from the Sag-Carina arm at $\sim 2$\,kpc, also shows a dip in the integrated HI profile at this velocity.  Thus all these features appear to be associated with near-distance solutions.

On the other hand, the weak feature at $-8$ km\,s$^{-1}$ that is also discussed in \S\ref{sec:pvplots} does not show any evidence for an HI absorption dip.  Given that the near-distance is only $\sim 1$\,kpc and so any absorption is unlikely to be filled in by foreground HI, this suggests that the far-distance solution ($\sim 13$\,kpc) is appropriate in this case.

\section{Conclusions}
\label{sec:conclusions}
We are undertaking a 3\,mm-band mapping survey of four isotopologues of the CO J=1--0 line from the southern Galactic plane using the CSIRO--ATNF Mopra radio telescope in Australia.  In this paper we present results from the first square degree of the survey ($l = 323 - 324^{\circ}, b = \pm 0.5^{\circ}$), providing a description of the methodology of the survey and the principal characteristics of the data set. The survey is being conducted using on-the-fly mapping with the telescope, and the final data cubes achieve a spatial and spectral resolution of $35''$ and $0.1$\,km/s, across $> 400$\,km/s bandwidth, in 4 lines ($\rm ^{12}CO, ^{13}CO, C^{18}O, C^{17}O$).  This provides a significant improvement in spatial and spectral resolution and in bandpass over previous CO surveys conducted of the $4^{th}$ quadrant of the Galaxy.  It results in a comprehensive data set in two lines ($\rm ^{12}CO$ and $\rm ^{13}CO$), with the $\rm C^{18}$O line detected at the brightest locations.  The sensitivity is not sufficient to detect the $\rm C^{17}O$ line.

We discuss the data reduction process, in particular the additional routines needed to remove bad pixels, rows and columns, and to baseline the data, following the pipeline processing through the Observatory standard mapping reduction software ({\sc livedata} \& {\sc gridzilla}).  We present images quantifying the beam coverage, system temperature and noise levels in the data set.  The sensitivities achieved are $\sim 1.5$\,K for $\rm ^{12}CO$ and $\sim 0.7$\,K for the other three lines, per 0.1\,km/s velocity channel.

We present diagnostic plots to aid in interpretation of the CO data set, in particular relating the $\rm ^{12}CO$ and $\rm ^{13}CO$ line brightness and ratio to the optical depth, $\rm [^{12}C/^{13}C]$ isotope ratio and gas excitation temperature.

We show a variety of sample spectra and images to illustrate the type and quality of data being obtained.  CO line emission is spread across a few tens of spectral features at negative $V_{LSR}$ velocities in this portion of the Galaxy. Peak line brightnesses for $\rm ^{12}CO$ are generally in the range $T_A^* \sim 1-5$\,K along sightlines across the survey field but rises to $\sim 6 - 13$\,K through the emission peaks.   We calculate line ratios, optical depths, column densities and molecular masses in several apertures chosen to encompass some of the emission features identified.  We also examine these quantities as a function of distance from the Sun, under the assumption that the $V_{LSR}$ velocities can be associated with the near-distance of the Galactic rotation curve.  We justify this assumption through comparison with a model for the Galaxy as well as by examination of HI self-absorption with $^{13}$CO emission features. Both line ratios and optical depths are found to be reasonably constant with distance (though with some significant variation in magnitude), with mean values of $\rm ^{12}CO/^{13}CO \sim 7$ and $\rm \tau_{12} \sim 9$.  Column densities of H$_2$ for the brightest clouds encountered along the sight lines are typically $\rm \sim 10^{22} cm^{-2} \equiv A_v \sim 10$\,mags.  Peak column densities along the sightlines through the brightest pixels are an order of magnitude higher.  Masses of the brightest clouds are of $\rm 10^{4} M_{\odot}$.  The total molecular mass within the 1 square degree region surveyed is $\rm \sim 2 \times 10^6 M_{\odot}$. Typical mass densities for the molecular clouds are $\rm \sim 0.1 \, M_{\odot} \, pc^{-3}$, when averaged over this survey area. 

Column densities are calculated using an empirical X--factor ($\rm X_{CO}$) from \citet{2001ApJ...547..792D} as well as via a radiative transfer calculation using the derived optical depth.  For the typical values found for $\tau_{12}$ the two methods are roughly equivalent when  $\rm [CO/H_2] \sim 3 \times 10^{-5}$ and $T_{ex} \sim 10$\,K\@.

This survey is ongoing, with our intention being to cover $l = 305 - 345^{\circ}, b = \pm 0.5^{\circ}$ when completed in 2015.  The data set will be also be made publicly available via the ATNF online archive.\footnote{atoa.atnf.csiro.au.}  It may also be accessed from the Mopra CO survey website\footnote{www.phys.unsw.edu.au/mopraco.}.  An additional CO data set from the Central Molecular Zone has also been obtained and will be published elsewhere.\footnote{See also www.phys.unsw.edu.au/mopracmz.}



\section*{Acknowledgments} 
The Mopra radio telescope is part of the Australia Telescope National Facility which is funded by the Commonwealth of Australia for operation as a National Facility managed by CSIRO\@. Many staff of the ATNF have contributed to the success of the remote operations at Mopra.  We particularly wish to acknowledge the contributions of David Brodrick, Philip Edwards, Brett Hisock, Balt Indermuehle and Peter Mirtschin. The University of New South Wales Digital Filter Bank used for the observations with the Mopra Telescope (the UNSW--MOPS)  was provided with support from the Australian Research Council (ARC). We also acknowledge ARC support through Discovery Project DP120101585.  This work was also carried out, in part, at the Jet Propulsion Laboratory, California Institute of Technology.   Finally, we thank the anonymous referee whose comments have helped improved this paper.

\clearpage

\begin{table*}[h]
\begin{center}
\caption{Parameters for the CO Line Observations with Mopra}
\label{tab:lineparams}
\begin{tabular}{ccccccccc}
\hline 
{\bf Line$^a$} & {\bf Rest Freq.} & {\bf IF$^b$} & ${\bf V_{\rm \bf low}}^c$ & ${\bf V_{\rm \bf high}}^c$  & {\bf Channel} & {\bf Number of} & ${\bf \sigma_{\rm \bf cont}}^d$ & ${\bf  T_{sys}}^e$ \\
&  & {\bf Number} & & & {\bf Spacing}  & {\bf Channels} &&  \\
& GHz & $\rm km\,s^{-1}$ & $\rm km\,s^{-1}$ & $\rm km\,s^{-1}$ & $\rm km\,s^{-1}$ & & K & K \\
\hline 
$\rm ^{13}CO$ J=1--0 & 110.201353 & 1 & $-135$ & $+235$ & 0.0916 & 4,096 & 1.0 & 420 \\
&& 1, 2 & $-500$ & $+240$ & 0.0916 & 8,192 & 0.7 & 450 \\
$\rm C^{18}O$ J=1--0 & 109.782173 & 2 & $-150$ & $+220$ & 0.0920 & 4,096 & 0.9 & 420 \\
&& 3, 4 & $-500$ & $+230$ & 0.0920 & 8,192 & 0.7 & 450 \\
$\rm C^{17}O$ J=1--0 & 112.358985 & 3 & $-265$ & $+95$ & 0.0899 & 4,096 & 0.9 & 470 \\
&& 5 & $-260$ & $+100$ & 0.0899 & 4,096 & 0.8 & 500 \\
$\rm ^{12}CO$ J=1--0 & 115.271202 & 4 & $-225$ & $+130$ & 0.0876 & 4,096 & 1.8 & 800 \\
&& 6, 7, 8 & $-550$ & $+500$ & 0.0876 & 12,288 & 1.5 & 810 \\
\hline
\end{tabular}
\medskip\\
\end{center}
$^a$For the spectral configurations (columns 3--7) for each line the first row refers to the configuration used in 2011 and the second to 2012 (when 4 additional zoom bands were added).\\
$^b$Reference numbers used for the zoom bands (IF $\equiv$ `Intermediate Frequency').\\
$^c$Approximate velocity limits ($V_{LSR}$) for data cubes; precise limits depend on the date of observation since the central frequency is fixed. \\
$^d$For each spectral line two values are shown.  The first row lists the standard deviation in the continuum channels over all pixels and the second row the mode value of their distribution (i.e.\ the 1$\sigma$ noise); see also Fig.~\ref{fig:sigma} for the distribution between pixels.\\
$^e$For each spectral line two values are shown. The first row lists median values for the system temperature across all pixels and the second row their mode values  (see also Fig.~\ref{fig:tsyshist} for the distributions).\\
\end{table*}

\begin{table*}[h]
\begin{center}
\caption{Parameters for the Fast Mapping of One Footprint with Mopra}
\label{tab:scanparams}
\begin{tabular}{lc}
\hline
{\bf Parameter} & {\bf Value} \\
\hline
Scan length & $60'$ \\
Scan width & $6'$ \\
Cycles per scan & 50 \\
Scans per Footprint & 24 \\
Bins per cycle & 8 \\
Cycle time & 2.048\,s \\
Bin time & 256\,ms \\
Scan rate & $35''/$\,s \\
Cell size (in scan direction) & $9''$ \\
Separation between scans & $15''$ \\
Number of scans per reference & 1 \\
Cycles per reference & 7 \\
Time between ambient temperature loads & 25\,min \\
Pixel size in final cube & $30''$ \\
Clock time per footprint & $\sim 1$\,hr \\
\hline
\end{tabular}
\medskip\\
\end{center}
Each $1^{\circ} \times 1^{\circ}$ survey region is scanned in both galactic longitude ($l$) and latitude ($b$), requiring a total of $10+10=20$ footprints (each of size $60' \times 6'$) to complete.
\end{table*}

\begin{table*}[h]
\begin{center}
\caption{Selected Apertures for the Analysis}
\label{tab:apertures}
\begin{tabular}{cccccccc}
\hline
{\bf Aperture} & $\bf l_{start}$ & $\bf l_{end}$ & $\bf b_{start}$ & $\bf b_{end}$ & $\bf V_{start}$ & $\bf V_{end}$ & {\bf Angular Size} \\
& deg. & deg. & deg. & deg. & km/s & km/s & sq. deg.\\
\hline
All & 323.00 & 324.00 & $-0.50$ & $+0.50$ & $-100$ & 0 & 1.0 \\
A   & 323.36 & 323.65 & $-0.11$ & $+0.12$ &   $-68$ & $-61$ & 6.7 (-2) \\
B   & 323.44 & 323.67 & $+0.25$ & $+0.40$ &   $-90$ &  $-78$ & 3.5 (-2) \\
C   & 323.76 & 323.97 & $-0.07$ & $+0.12$ &   $-65$ &  $-45$ & 4.0 (-2) \\
D   & 323.00 & 323.20 & $-0.46$ & $-0.26$ & $-37$ & $-27$ & 4.0 (-2)\\
E   & 323.35 & 323.48 & $-0.26$ & $-0.19$ &   $-68$ & $-61$ & 9.1 (-3) \\
F   & 323.07 & 323.21 & $+0.13$ & $+0.26$ &   $-68$ & $-61$ & 1.8 (-2) \\
\hline
\end{tabular}
\medskip\\
\end{center}
Coordinates for the apertures selected for further analysis.  These are also marked on Fig.~\ref{fig:apertures}.
\end{table*}

\begin{table*}[h]
\begin{center}
\caption{Adopted Parameters for Selected Apertures}
\label{tab:apertureparams}
\begin{tabular}{ccccccc}
\hline
{\bf Aperture} &  {$\bf V_{\bf LSR}$} & {\bf Distance} & {\bf Size} & {\bf Area} & {\bf Scale Size$^c$} & ${\bf [^{12}C/^{13}C]}$  \\
& km/s & kpc &  pc & pc$^2$ & (for $30''$) pc & {\bf Ratio} \\ \hline
All & N/A$^a$  & N/A & N/A & N/A & N/A & 57 \\
A   & $-65$ & 4.8 & 22 & 460 & 0.7 & 54 \\
B   & $-84^b$ & 6.8 & 22 & 490 & 1.0 & 51 \\
C   & $-55$ & 4.1 & 14 & 200 & 0.6 & 56 \\
D   & $-32$ & 2.5 &  8.7 & 76  & 0.4 & 61 \\
E   & $-65$ & 4.8 &  7.9 & 63  & 0.7 & 54 \\
F   & $-65$ & 4.8 &  11 & 130  & 0.7 & 54 \\
\hline
\end{tabular}
\medskip\\
\end{center}
Parameters adopted for the apertures listed in Table~\ref{tab:apertures}, as described in \S\ref{sec:apertures}. The distance is the near-value for the $V_{LSR}$ radial velocity.  Size refers to the areal size of the aperture at that distance.  \\
$^a$ For the integrated aperture (`All') no sensible velocity, and hence distance or sizes, can be defined as this includes emission from the entire sight line. \\
$^b$ ``Forbidden'' velocity, so the distance is set to that of the tangent point at $l = 323.5^{\circ}$.\\
$^c$ Scale size for a $30''$ beam at the source distance.
\end{table*}


\begin{table*}[h]
\begin{center}
\caption{Line Fluxes for Selected Apertures}
\label{tab:linefluxes}
\begin{tabular}{cccccccccc}
\hline
 {\bf Aperture} & \multicolumn{4}{c}{\bf Integrated Line Flux} & \multicolumn{2}{c}{\bf Peak Brightness} & \multicolumn{2}{c}{\bf Peak Position} & {\bf Dame et al.}\\
& $^{12}$CO & $^{13}$CO & C$^{18}$O & C$^{17}$O & $^{12}$CO & $^{13}$CO & $l$ & $b$ & $^{12}$CO \\
& \multicolumn{4}{c}{K km/s} & K & K & deg. & deg. & K km/s\\  
\hline
All & 81.0 & 11.0 & 1.4 & 0.6 & 13.1 & 4.7 & 323.45 & +0.08 & 62.6 \\
     &   3.1 &   1.5 & 1.4 & 2.0 \\
A   & 29.7 &   5.7 & 1.1 & 0.1 & 13.1 & 4.7 & 323.45 & +0.08 & 19.5 \\
     &   0.6 &   0.3 & 0.3 & 0.4 \\
B   & 21.0 &   3.2 & 1.5 & -0.5 & 6.1 & 2.3 & 323.57 & +0.36 & 11.9 \\
     &   1.3 &   0.6 & 1.2 & 0.8 \\
C   & 52.2 &   7.4 & 2.1 & 1.1 & 11.7 & 4.0 & 323.90 & $+0.03$ & 42.1 \\
     &   2.1 &   1.2 & 1.1 & 1.1  \\
D   & 29.2 &   4.4 & 0.2 & 0.6 & 8.1 & 3.7 & 323.05 & $-0.42$ & 20.6 \\
     &  1.0  &   0.5 & 0.5 & 0.6 \\
E   & 21.9 &  4.2 & 0.6 & -0.2 & 12.6 & 4.1 & 323.44 & $-0.19$ & 10.4 \\
    &    1.5 &  0.7 & 0.8 & 0.9 \\
F  & 29.6  &  6.9 & 2.1 & 0.2 & 11.7 & 6.2 & 323.19 & +0.16 & 14.4 \\
   &   0.9  &  0.5  & 0.5 & 0.5 \\
\hline
\end{tabular}
\medskip\\
\end{center}
Columns 2--5 tabulate the integrated line fluxes and their errors (underneath) in K km/s, for the four isotopologues observed of CO J=1--0\@, in each of the apertures defined in Table~\ref{tab:apertures}.  They include a correction for the beam efficiency, $\eta_{XB}$, of 0.55.  Errors include both the statistical error as well as an estimate for the error in determining the continuum level. Columns 6--7 list the peak line channel brightness ($T_A^*$, in K) within each aperture for $^{12}$CO and $^{13}$CO\@.  Columns 8--9 list the corresponding spatial positions for the peak pixel. The final column [10] shows the line flux determined for the aperture from the \citet{2001ApJ...547..792D} $^{12}$CO data cube.
\end{table*}


\begin{table*}[h]
\begin{center}
\caption{Line Ratios and Optical Depths for Selected Apertures}
\label{tab:lineratios}
\begin{tabular}{ccc}
\hline
{\bf Aperture} & \multicolumn{2}{c}{\bf Parameter} \\
& $^{12}$CO/$^{13}$CO & $\tau_{12}$  \\
\hline
All & 7.4 & 7.7 \\
     & 1.3 & 1.3 \\
A   & 5.2 & 10.5 \\
     & 0.4 & 0.8 \\
B   & 6.6 & 7.8 \\
     & 1.6 & 1.9 \\
C   & 7.1 & 7.9 \\
     & 1.4 & 1.6 \\
D  & 6.6 & 9.2 \\
    & 1.0 & 1.3 \\
E  & 5.3 & 10.3 \\
    & 1.3 & 2.5 \\
F  & 4.3 & 12.7 \\
   & 0.4 & 1.2 \\
\hline
\end{tabular}
\medskip\\
\end{center}
The $^{12}$CO/$^{13}$CO J=1--0 line ratios and optical depths for the $^{12}$CO line, $\tau_{12}$, derived from the integrated fluxes for each of the apertures defined in Table~\ref{tab:apertures}, with their corresponding $1 \sigma$ errors listed underneath.  The optical depths are derived from the line ratios and the $\rm [^{12}C/^{13}C]$ isotope ratios listed in Table~\ref{tab:apertureparams}, when applying eqn.~\ref{eqn:opticaldepthapprox}.
\end{table*}

\begin{table*}[h]
\begin{center}
\caption{Column Densities and Masses for Selected Apertures}
\label{tab:columndensity}
\begin{tabular}{cccc}
\hline
{\bf Aperture} & \multicolumn{3}{c}{\bf Parameter} \\
& $\rm N_{H_2}$ / 10$^{21}$ cm$^{-2}$ & Mass / $10^4$ (M$_{\odot}$) & $\rm N_{H_2}$ / 10$^{21}$ cm$^{-2}$ \\
&  (X$_{\rm CO}$--factor) && (T = 10\,K +  $\tau_{12}$)    \\
\hline
All & 21.9 & 170$^{a}$ & 15.7 \\
     &   0.8 &        &   0.6 \\
A   &   8.0 &  6.9 &   7.8 \\
     &   0.2 &  0.2 &   0.2 \\
B   &   5.7 &  5.2 &   4.1 \\
     &   0.4 &  0.3 &   0.3 \\
C   & 14.1 &  5.3 &  10.3 \\
     &   0.6 &  0.2 &   0.4 \\
D   &  7.9 &  1.1 &   6.8 \\  
     &  0.3 & 0.04 &  0.2 \\
E   &  5.9 & 0.69 &  5.7 \\
     &  0.4 & 0.05 &  0.4 \\
F   &  8.0 & 1.9   &  9.4 \\
     &  0.2 & 0.1   &  0.3 \\
\hline
\end{tabular}
\medskip\\
\end{center}
The second column shows the column density of H$_2$ molecules derived from the integrated line flux for each of the apertures listed in the first column (as defined in Table~\ref{tab:apertures}).  This is calculated by applying an $\rm X_{CO}$ factor of $\rm 2.7 \times 10^{20} cm^{-2} (K\,km\,s^{-1})^{-1}$ (\citet{2001ApJ...547..792D} for $|b| < 1^{\circ}$).  The corresponding gas mass (including a contribution of 10\% Helium by H--atom number) is listed in the next column.   The last column gives the H$_2$ column density assuming a gas temperature of 10\,K, correcting for the optical depth listed in Table~\ref{tab:lineratios}, and using a [CO/H$_2$] abundance ratio of $3 \times 10^{-5}$ (i.e.\ as given by eqn~\ref{eqn:xfactor}). For gas temperatures of 20\,K, 40\,K and 80\,K these column densities should be multiplied by factors of 1.5, 2.7 and 4.9, respectively, as determined using the right hand plot in Fig.~\ref{fig:coanalysis}. In all cases the second row for each aperture gives the corresponding $1 \sigma$ error.  \\
$^{a}$ No mass can sensibly be given from this analysis for the integrated line flux over the entire survey region (i.e.\ aperture ``All'') as this encompasses emission from a wide range of distances. The number quoted here comes from the integrated mass distribution, as shown in Fig.~\ref{fig:radialcut}  and discussed in \S\ref{sec:massdistribution}.
\end{table*}

\clearpage

\clearpage

\clearpage


\end{document}